\documentclass[
reprint,prb,
superscriptaddress,
nofootinbib,
amsmath,amssymb,
aps,
]{revtex4-2}

\usepackage{amsmath}
\usepackage{amssymb}
\usepackage{braket}	
\usepackage{xcolor}
\usepackage{appendix}
\usepackage{dsfont}
\usepackage{graphicx}
\usepackage[unicode=true,pdfusetitle,
 bookmarks=false,bookmarksnumbered=false,bookmarksopen=false,
 breaklinks=false,pdfborder={0 0 1},backref=false,colorlinks=true,urlcolor=blue,citecolor=blue,linkcolor=black]
 {hyperref}

\newcommand{\Trace}[1]{\text{Tr}\left( #1\right)}

\newcommand{\tel}{\tau_\text{el}}
\newcommand{\e}{\varepsilon}
\newcommand{\ww}{\omega}
\newcommand{\qb}{\mathbf{k}}
\newcommand{\kb}{\mathbf{k}}
\newcommand{\rb}{\mathbf{r}}
\newcommand{\msf}[1]{\mathsf{#1}}

\newcommand{\gm}{\gamma_{\msf{m}}}
\newcommand{\gd}{\gamma_{\msf{D}}}

\newcommand{\vex}[1]{\mathbf{#1}}

\DeclareMathOperator{\Tr}{Tr}
\DeclareMathOperator{\tr}{tr}

\begin{document}

\title{
Measurement-induced phase transitions in disordered fermions
	}
\begin{abstract}
Measurement-induced phase transitions are nonequilibrium transitions between  phases characterized by distinct entanglement scaling behaviors, driven by the competition between unitary dynamics and measurements. Despite recent numerical efforts, how quenched disorder affects these transitions remains unclear. In this work,
we study a $d$-dimensional noninteracting fermionic system subject to both quenched disorder and continuous monitoring of the local particle density, and derive an effective field theory describing its long-time universal behaviors. We find that the system is governed by the same nonlinear sigma model as in the case of clean monitored fermions, with disorder entering only through a modification of model parameters. 
This result suggests that the presence or absence of a measurement-induced phase transition is unaffected by the introduction of disorder. In spatial dimension $d=1$, the system exhibits only an area law phase and no transition, whereas in
$d>1$, a transition persists between an area$\times$log law  phase and an area law phase, with the critical measurement rate suppressed by disorder. 
Numerical simulation confirms the absence of MIPT in $d=1$, and the linear shift of the critical measurement rate by disorder scattering rate in higher dimensions predicted by analytical theory.
\end{abstract}

\author{Yunxiang Liao}
\email{liao2@kth.se}
\affiliation{Department of Physics, KTH Royal Institute of Technology, SE-106 91
Stockholm, Sweden}

\author{Max Matheussen}
\affiliation{Department of Physics, KTH Royal Institute of Technology, SE-106 91
Stockholm, Sweden}
\affiliation{Department of Physics, Technical University of Denmark, 2800 Kgs.~Lyngby, Denmark}

\author{Xinghai Zhang}
\affiliation{Department of Physics, KTH Royal Institute of Technology, SE-106 91
Stockholm, Sweden}
    
\maketitle

\section{Introduction}


Measurement-induced phases and phase transitions (MIPTs) have attracted significant attention from both the condensed matter and quantum information communities in recent years. These phenomena not only deepen our fundamental understanding of information and entanglement dynamics driven by the competition between unitary scrambling and local measurements, but they are also highly relevant to quantum computation, as quantum error correction and coupling to the environment inevitably introduce effective measurement-like nonunitary dynamics.

MIPTs were first extensively investigated in random quantum circuits \cite{Skinner_MIPT_2019_Circuits,Fisher_QC_2018,Chan_QC_2019,Szyniszewsmi_QC_2019,Li_qc_2019,Bao_qc_2020,Jian_qc_2020,Zhang_qc_2020,Zabalo_qc_2020,Gullans_qc_2020,Gullans_qc_Oct_2020,Iaconis_qc_2020,Turkeshi_qc_2020,Szyniszekwi_qc_2020,Choi_QC_2020,Li_qc_2021,Sang_qc_2021,Nahum_qc_2021,Ippoliti_qc_2021,Lavasani_qc_2021,Lavasani_qc_Dec_2021,Li_qc_2021,Block_qc_2022,Agrawal_qc_2022,Barratt_qc_2022,Shradda_qc_2022,Kelly_qc_2023,Morral_qc_2023,Zabalo_qc_2023,Fisher_qc_review_2023,Ha_qc_2024,Liu_qc_2024,manna_qc_2024,Majidy_qc_2024,Agrawal_qc_2024,Chakraborty_qc_2024,Yang_qc_2025,Khanna_qc_2026}, where a transition between volume-law and area-law entanglement phases occurs at a critical measurement rate. More recently, this research has been extended to various other platforms, including complex fermions with $\mathrm{U}(1)$ charge conservation \cite{Cao_free_1D_numerical_2019,Xiao_1D_free_fermions_2020,Diehl_free_1D_numerical_2021,Diehl_free_1D_analytical_2021,Turkeshi_free_fermion_1D_2021,Tang_2D_free_fermions_2021,Turkeshi_free_fermions_1D_2022,Zhang_free_fermions_1D_2022,Carollo_free_fermions_2022,Mianto_fermion_2022,Buchhold_free_fermions_1D_2022,Coppola_1D_free_fermions_2022,Mirlin_free_1D_2023,Nahum_free_1D_majorana_and_dirac_2023,Gal_free_fermions_1D_2023,Buchhold_free_fermions_dark_state_2023,Romito_free_fermions_2023,Nahum_fermions_free_2024,Piccitto_free_fermion_2024,Buchhold_free_fermions_1D_2D_2025,Sieberer_free_fermions_2025,Fan_free_fermions_numerical_2026,Mirlin-fluctuations,Mirlin-Levy,Mirlin-periodic}, Majorana fermions \cite{Jian_free_Majorana_qc_1D_2023,Turkeshi_free_majorana_2023, Nahum_free_1D_majorana_and_dirac_2023,foster2025Maj},
free bosons \cite{Yokomizo2025}, and disordered systems \cite{Popperl_AL_Meas_2023, Szyniszewski_Disordered_free_fermions_2023,Yamamoto_disorder_MBL_Bosons_2023,Khaymovich_MBL_bosons_2024,Zhao_disordered_free_fermion_1D_2026}.
For complex fermions in one dimension, the field was initially divided, with a group of investigations concluding that there existed a phase transition from a logarithmic phase, scaling with $\log{\ell}$, where $\ell$ denotes the size of a subsystem, to the area law phase \cite{Diehl_free_1D_analytical_2021,Turkeshi_free_fermion_1D_2021,Mianto_fermion_2022}, and another stating that no genuine phase transition existed \cite{Cao_free_1D_numerical_2019,Coppola_1D_free_fermions_2022,Sieberer_free_fermions_2025}.
It was later discovered that monitored free fermions can be mapped to a Keldysh nonlinear sigma model (NL$\sigma$M) \cite{Mirlin_free_1D_2023, Nahum_free_1D_majorana_and_dirac_2023}. This NL$\sigma$M has the same structure as the one governing the Anderson localization transition \cite{Anderson_Transition_Review}, but is evaluated at a different replica limit ($R \to 1$ instead of $R \to 0$). A renormalization group (RG) analysis of this NL$\sigma$M reveals that no genuine phase transition for 1D free fermions survives in the asymptotic limit $\ell \to \infty$~\cite{Mirlin_free_1D_2023}. Instead, the intermediate scale logarithmic entanglement entropy  actually saturates at the correlation length, where it gives way to strictly area-law entanglement. This analytical prediction was further corroborated by careful numerical calculations reaching much larger system sizes \cite{Fan_free_fermions_numerical_2026}. Further analytical and numerical studies confirm that the MIPT of free complex fermions only occurs in higher dimensions ($d\ge 2$),
where a phase transition exists between a area$\times$log law $\ell^{d-1}\log{\ell}$, and the area law phase \cite{Buchhold_free_2D_2024,Mirlin_free_2D_2024,Jin2024,Fan_free_fermions_numerical_2026}. 

Realistic physical systems are rarely completely clean; rather, ubiquitous quenched disorder significantly alters the wave functions of fermionic systems, driving Anderson localization in low dimensions or at strong disorder strengths \cite{Anderson_localisation_original, Anderson_Transition_Review}. 
In static free fermion systems with quenched disorder, it is well established that the entanglement entropy obeys an area law, even in the weak disorder delocalized phase \cite{burmistrov2017entanglement}. However, previous studies have shown that local measurements can introduce delocalization corrections for dirty fermions \cite{Popperl_AL_Meas_2023}, and even stabilize the critical phase and the associated MIPT in 1D free fermions \cite{Szyniszewski_Disordered_free_fermions_2023}. Therefore, it is very interesting to further investigate the interplay between measurement and disorder to determine the ultimate fate of MIPTs in fermionic systems.
Additionally, the quantum corrections responsible for the absence (presence) of an MIPT for monitored fermions in $d=1$ ($d>1$) are similar to the weak localization corrections in static disordered systems, with both described by analogous NL$\sigma$M. Disordered and monitored systems thus provides a rich arena to explore these closely related but phenomenologically different effects.

In this work, we develop a two-matrix field theory within the replica Keldysh formalism to describe a monitored disordered system of noninteracting fermions, with a local-in-time matrix field $\hat{q}_{t}$ capturing the effect of continuous monitoring, and a bilocal-in-time matrix field $\hat{Q}_{tt'}$ capturing the effect of disorder. We find that the Goldstone modes, governing the universal long-time, long-distance behavior of the system, is generated by applying the same local-in-time unitary rotation in replica and Keldysh spaces to the saddle points of both fields.
The allowed local-in-time rotations are restricted to the same subset that generates soft modes in the measurement-only (clean monitored) theory. 
Consequently, the Goldstone fluctuations are described by an NL$\sigma$M with the same structure 
and the same target manifold as in the measurement-only theory, but with parameters modified by disorder. Specifically, the stiffness $\lambda$ changes from $\lambda \propto 1/\gm$ in the measurement-only theory to $\lambda \propto 1/\tilde{\gamma}$, where $\tilde{\gamma}=\gm+\gd$ is the total decay rate including both the disorder-induced elastic scattering rate $\gd$ and the measurement-induced decay rate $\gm$.

By contrast, the remaining fluctuations that are massless in the disorder-only or measurement-only theory now become massive in the presence of both disorder and measurement, and can be separated into three categories with different origins. First, fluctuations generated by applying the same local-in-time rotation to both $\hat{q}$ and $\hat{Q}$ are massive if that rotation do not belong to the measurement-only Goldstone manifold.
Second, relative local-in-time rotations of $\hat{q}$ and $\hat{Q}$, in which the two fields are rotated by different local-in-time unitaries, also acquire a finite mass once measurement and disorder are both present. These first two types of massive modes are local in time, and when integrated out lead to a renormalization of NL$\sigma$M parameters. In particular, they generate the factor $z=\tilde{\gamma}/2\gm$ that renormalizes the temporal gradient and source terms in the replica-asymmetric sector of the NL$\sigma$M.
Third, a separate class of massive fluctuations originates from nonlocal-in-time rotations of $\hat{Q}$, which are soft in the disorder-only theory but become massive in the presence of measurement. At the quadratic order, they decouple from the local-in-time Goldstone modes, and can be ignored away from the limit  $\gm/\gd \ll 1$ where their mass becomes small. 
As a result,  in the regime of weak disorder and measurement strengths (when both $\gm,\gd$ are much smaller than the characteristic kinetic energy scale), and away from the disorder dominated limit $\gm/\gd \ll 1$, the low-energy dynamics is governed entirely by the local-in-time Goldstone modes. The resulting NL$\sigma$M has the same form as in the measurement-only theory, with disorder only modifying its parameters.

The fact that the quenched disorder leave the form of the NL$\sigma$M unchanged and enter only through the modification of parameters imply that it does not affect the presence or absence of a measurement-induced phase transition.
In $d=1$, the measurement-induced phase transition remains absent, whereas for $d>1$ a transition between critical and area-law phases persists, but with a shifted critical measurement rate. In particular, since the introduction of  disorder changes the stiffness of the NL$\sigma$M from $1/\gm$ to $1/(\gm+\gd)$, it suppresses the critical measurement rate in $d>1$. 
Our numerical simulations support these sigma-model predictions. Specifically,
our 1D results demonstrate that disorder does not stabilize the MIPT in 1D. Instead, there is no genuine MIPT in disordered 1D fermions, in contrast to previous numerical claims based on small system sizes~\cite{Szyniszewski_Disordered_free_fermions_2023,Zhao_disordered_free_fermion_1D_2026}. Our 2D results confirm that the MIPT persists in the presence of disorder potential and that the critical measurement rate $\gm^c$  decreases linearly with increasing $\gd$, consistent with the analytical prediction.


The rest of the paper is organized as follows. In Sec.~\ref{sec:background}, we introduce the theoretical background and formulate the replica Keldysh framework for the monitored disordered fermions. In Sec.~\ref{sec:2field}, we derive a two-matrix field theory and perform a saddle-point analysis. In Sec.~\ref{Sec:Gau}, we investigate Gaussian fluctuations around the saddle point configuration in both the replica-symmetric and replica-asymmetric sectors and evaluate the averaged density response function as well as the second cumulant of the number of particles in a subsystem, from which the entanglement entropy can be extracted within the Gaussian approximation. In Sec.~\ref{sec:NLSM}, we derive the NL$\sigma$M that describes the Goldstone modes governing the long-time behavior of entanglement entropy and charge fluctuations in the weak measurement and weak disorder regime. We show that this effective theory coincides with that of the measurement-only case after a replacement of parameters. 
In Sec.~\ref{sec:num}, we present numerical evidence supporting our analytical findings, demonstrating the absence (presence) of an MIPT in 1D (2D), as in the clean case. We further show that in 2D, disorder shifts the MIPT to a lower critical measurement strength, with the critical $\gm$ decreasing linearly with increasing $\gd$.  Finally, in Sec.~\ref{sec:conclusion}, we summarize the main results and discuss possible future directions. Additional technical details are provided in the Appendices. Appendix~\ref{sec:app-dp} presents the explicit expressions for the dressed propagators of local-in-time Gaussian fluctuations, while Appendix~\ref{sec:app-dp-2} extends the analysis to include massive bilocal-in-time fluctuations. 
Further details of the NL$\sigma$M derivation are provided in Appendix~\ref{sec:app-nlsm}, while the numerical approach is discussed in Appendix~\ref{sec:app-numerics}.

\section{Model and Replica Keldysh Framework}\label{sec:background}

We consider a $d$-dimensional system of noninteracting fermions subject to both quenched disorder and continuous monitoring of the local particle density $n_{\rb}$.
The dynamics over a time interval $\delta t$ consists of (1) the unitary evolution $e^{-i\hat{H}\delta t}$ generated by the Hamiltonian $\hat{H}=\hat{H}_0+\hat{H}_{\msf{dis}}$, where $\hat{H}_0$ represents the hopping between different sites and $\hat{H}_{\msf{dis}}$ denotes a static on-site disorder potential, and (2) the weak measurement of the local particle density described by a set of Kraus operators:
\begin{align}
    \hat K_{m_\mathbf{r}}=
    \mathcal{N}_{m_\mathbf{r}}
    \exp \left( -\frac{\kappa \delta t}{2}(\hat n_{\mathbf{r}}-m_\mathbf{r})^2 \right) .
\end{align}
Here $\kappa$ denotes the measurement strength, 
$m_\mathbf{r}$ is the random  measurement outcome at site $\rb$. The probability of observing a given outcome is determined by the Born rule~\cite{nielsen2010quantum}
$
    P_{m_\mathbf{r}}=\Tr
    \left(
    \hat K_{m_\mathbf{r}}
    \hat{\rho}
    \hat K_{m_\mathbf{r}}^{\dagger}
    \right),
$
where $\hat{\rho}$ is the density matrix right before the measurement.
The normalization constant $\mathcal{N}_{m_\mathbf{r}}$ is chosen to meet the positive-operator valued
measure (POVM) condition,
$
    \sum_{m_{\rb}}
    \hat K_{m_{\mathbf{r}}}^{\dagger}
    \hat K_{m_{\mathbf{r}}}
    =
    \mathds{1}.
$
Combining the unitary time evolution and measurement, the density matrix $\hat{\rho}$ evolves according to
\begin{align}
\begin{aligned}
    &\hat{\rho} (t+\delta t)
    =
    \frac{\hat{\tilde{\rho}} (t+\delta t)}{\Tr\left[\hat{\tilde{\rho}}(t+\delta t)\right]},
    \\
    &
    \hat{\tilde{\rho}} (t+\delta t)=
    \left(\prod_{\rb}\hat{K}_{m_{\mathbf{r},t}}\right)  e^{-i\hat{H}\delta t} \hat{\rho}(t) e^{+i\hat{H}\delta t}\left(\prod_{\rb}\hat{K}^{\dagger}_{m_{\mathbf{r},t}}\right),
\end{aligned}
\end{align}
where $\hat{\tilde{\rho}}$ denotes the unnormalized post-measurement density matrix, and $m_{\mathbf{r},t}$ represents the measurement outcome at site $\rb$ and time $t$.


In the continuum limit, 
the long-time behavior of 
the system, after averaging over quantum trajectories and disorder realizations,
is described by the following replica Keldysh path integral 
\begin{align}\label{eq:Z}
    \begin{aligned} 
    Z[J]=&\overline{ \int\mathcal{D} (\bar{\psi},\psi)
    e^{iS_0+iS_V+iS_J} },
    \\
S_0=&\int_{t,t^\prime,\mathbf{r},\mathbf{r}^\prime}\sum_{j=1}^R\bar\psi_{\mathbf{r},t}^{j}\hat G_0^{-1}(\mathbf{r},\mathbf{r}^\prime;t,t^\prime)\psi_{\mathbf{r}^\prime,t^\prime}^{j},
    \\
    S_{V}=&
    \int_{t,\mathbf{r}}\sum_{j=1}^R 
    (-V_\text{dis}(\mathbf{r})n_{\mathbf{r},t}^{\text{q},j} + i V_\text{meas}(\mathbf{r},t)n^{\text{cl},j}_{\mathbf{r},t}),
    \\
    S_{J}
    =&
    -\int_{t,\mathbf{r}}\sum_{j=1}^{R}\left(J^{\text{q},j}_{\mathbf{r},t}n_{\mathbf{r},t}^{\text{cl},j}+J^{\text{cl},j}_{\mathbf{r},t}n_{\mathbf{r},t}^{\text{q},j}\right).
    \end{aligned}
\end{align}
Here $\psi_{\mathbf{r},t}^{a,j}$ carries both a replica index $j=1,2,...R$, together with a Keldysh index $a=+(-)$ corresponding to the forward (backward) Keldysh branch.
$n^{\text{cl}/q}$ is the classical/quantum component of the particle density, and is defined as
\begin{align}
\begin{aligned}
    n^{\text{cl}/q}_{\mathbf{r},t}=n^+_{\mathbf{r},t} \pm n^-_{\mathbf{r},t},
    \qquad
    n^{a}_{\mathbf{r},t}=\bar\psi^{a}_{\mathbf{r},t}\psi^{a}_{\mathbf{r},t},
    \qquad
    a=\pm.
\end{aligned}
\end{align}
We have used the short-hand notation $\int_{\mathbf{r}}= \int d^d \rb$ and $\int_{t}=\int_{-\infty}^{t_f} dt$, where the final time is set to $t_f=0$, by which the system has reached the dynamical steady state whose properties are studied.
 
The action $S_0$ describes the  unitary time evolution of noninteracting fermions in the absence of disorder and measurement. The corresponding bare Green's function $G_0$ is given by:
$
    G_0^{-1}
    =i\partial_t -\xi(-i\nabla),
$
where $\xi(k)$ denotes the single-particle kinetic energy. 
The action $S_{V}$ accounts for the effects of the static disorder potential $V_\text{dis}(\rb)$ and the continuous monitoring encoded by the effective measurement noise $V_\text{meas}(\rb,t)$, which can be viewed as the continuum limit of measurement outcomes.
Relative to previous field-theoretical studies of clean monitored fermions \cite{Mirlin_free_1D_2023,Mirlin_free_2D_2024,Buchhold_free_2D_2024}, the essential new ingredient of the present theory is the inclusion of the static disorder potential $V_\text{dis}(\rb)$.
Both  $V_\text{dis}(\rb)$ and  $V_\text{meas}(\rb,t)$ are taken to be Gaussian with zero mean and variances
\begin{align}
\begin{aligned}
    &\overline{V_\text{dis}(\rb)V_\text{dis}(\rb')}
    =
    \delta(\rb-\rb')/(2\pi \nu \tau_{el}),
    \\
    &\overline{V_\text{meas}(\rb,t) V_\text{meas}(\rb',t')}
    =g^2\delta(\rb-\rb')\delta(t-t'),
    \label{eqn:Gaussian}
\end{aligned}
\end{align}
where $g^2\propto\kappa$ characterizes the measurement strength, $\tau_{el}$ represents the elastic scattering time due to static impurities, and $\nu$ is the density of states of the underlying (unmonitored) system.
The overline here, and in the first line of Eq.~\ref{eq:Z}, represents averaging over  $V_\text{dis}(\rb)$ and $V_\text{meas}(\rb,t)$.


We also introduce a source term $S_J$ to study the full counting statistics of particle number and the entanglement properties. By taking the functional derivatives of $Z[J]$ in Eq.~\ref{eq:Z} with respect to the source field $J$, whose quantum (classical) component $J^{\msf{q/cl}}$ couples to the classical (quantum) component of local particle density $n^{\msf{cl/q}}$, we obtain correlation functions of the particle density and, consequently, the particle number cumulants.

For our system prepared in a Gaussian initial state, the particle number cumulants determine the von Neumann entanglement entropy $S_A$ between a subsystem $A$ and the rest of the system via the Klich and Levitov formalism~\cite{Klich-Levitov}:
\begin{equation}
    S_{A}=-\Tr (\hat{\rho}_A \ln \hat{\rho}_A)=\sum_{q=1}^\infty 2\zeta(2q)C_{A}^{(2q)}.
    \label{eqn:Klich-Levitov}
\end{equation}
Here $\hat{\rho}_A$ is the reduced density matrix for subsystem A, and $C_{A}^{(m)}$ is the $m$-th cumulant of the total particle number in this subsystem, $\hat{N}_A=\sum_{\mathbf{r}\in A}\hat{n}_\mathbf{r}$.
The expansion in Eq.~\ref{eqn:Klich-Levitov} is dominated by the leading term $S_{A}\approx \frac{\pi^2}{3} C_A^{(2)}$, in monitored free fermionic systems with weak measurement strength~\cite{Mirlin_free_1D_2023,Mirlin_free_2D_2024} and also in weakly disordered systems~\cite{burmistrov2017entanglement}.

Since this relation between the entanglement entropy $S_A$ and the particle-number cumulants $C_A^{(m)}$ holds for individual quantum measurement trajectories and fixed impurity potential configurations, it directly carries over to quantities averaged over both quantum trajectories and disorder realizations — the primary focus of this work. 
The quantum trajectory averaging can be performed using the replica trick.
In particular, the averaged $n$-replica of the normalized density matrix, required for the evaluation of the averaged $n$-th particle-number cumulant, can be obtained from
\begin{align}
\begin{aligned}
     \overline{\hat{\rho}^{\otimes n}}
     = \sum_{\gamma} \Tr^{1-n}(\hat{\tilde{\rho}}_{\gamma})\hat{\tilde{\rho}}^{\otimes n}_{\gamma}
    =\lim_{R\rightarrow 1} \text{Tr}_{r=n+1,\dots,R} \sum_{\gamma}
    \otimes_{r=1}^{R}\hat{\tilde{\rho}}_{\gamma}^{(r)}.
\end{aligned}    
\end{align}
Here the overline represents averaging over quantum trajectories labeled by subscript $\gamma$, each corresponding to a specific sequence of measurement outcomes $\left\lbrace m_{\rb,t} \right\rbrace$ and weighted by its Born rule probability. This Born rule weighting generates an additional factor of $\Tr(\hat{\tilde{\rho}}_{\gamma})$ in the first equality.
In the last equality, the first $n$ replicas generate $\hat{\tilde{\rho}}^{\otimes n}$, while tracing over the remaining $R-n$ replicas produces $\Tr^{R-n}(\hat{\tilde{\rho}})$. Analytic continuation to the $R \rightarrow 1$ limit then reproduces the quantum trajectory averaged $n$-replica normalized density matrix. While the replica trick is also widely used for quenched disorder averaging, involving $R \rightarrow 0$ as opposed to $R \rightarrow 1$ limit, in our framework
disorder averaging is handled directly within this replicated Keldysh framework and requires no additional replica structure.

Starting from Eq.~\ref{eq:Z},  we then apply the standard Larkin-Ovchinnikov rotation
\begin{align}
\begin{aligned}
    \psi \rightarrow \hat{\tau}^3 U_{\msf{LO}} \psi,
    \quad
    \bar{\psi} \rightarrow \bar{\psi} U_{\msf{LO}}^{\dagger},
    \quad
    U_{\msf{LO}}=(1+i\hat{\tau}^2)/\sqrt{2},
    \label{eqn:LO}
\end{aligned}
\end{align}
with $\hat{\tau}$ being the  Pauli matrix in the Keldysh space.
After the transformation, the bare Green's function in the Keldysh space becomes
\begin{equation}
    G_0^{-1}(\omega,\mathbf{k})=\begin{pmatrix}
        \omega-\xi_{\qb} +i\eta& 2i\eta F_0(\omega)\\
        0&\omega-\xi_{\qb}-i\eta
    \end{pmatrix}.
    \label{eqn:G0}
\end{equation}
$F_0(\omega)$ is the initial fermionic Keldysh distribution function and $\eta \rightarrow 0^+$ is a positive infinitesimal.
The rotation also brings
$n^{\text{cl}/q}=n^+ \pm n^-$ to the forms
\begin{equation}
    n^{\text{cl},j}_{\mathbf{r},t}=\bar\psi^{j}_{\mathbf{r},t}\hat{\tau}^1\psi^{j}_{\mathbf{r},t},\quad n^{\text{q},j}_{\mathbf{r},t}=\bar\psi^{j}_{\mathbf{r},t}\psi^{j}_{\mathbf{r},t}.
    \label{eqn:site occupation numbers in Keldysh space}
\end{equation}

\section{Effective two-matrix field theory}~\label{sec:2field}

\subsection{Hubbard-Stratonovich transformation}

The averaging over quantum trajectories and disorder realizations is carried out by averaging over $V_\text{meas}(\rb,t)$ and $V_\text{dis}(\rb)$, respectively, both of which are Gaussian distributed with the variances given by Eq.~\ref{eqn:Gaussian}.
This leads to two effective interactions with different time structures:
\begin{equation}
\begin{aligned}
    &iS_{\msf{ave}} =
    \ln \left( \overline{e^{iS_V}}\right),
    \\
    =
    & -\frac{1}{4\pi\nu\tel}\int_{t,t^\prime,\mathbf{r}}\sum_{i,j=1}^Rn_{\mathbf{r},t}^{\text{q},i} n_{\mathbf{r},t'}^{\text{q},j}
    +\frac{g^2}{2}\int_{t,\mathbf{r}}\sum_{i,j=1}^Rn_{\mathbf{r},t}^{\text{cl},i}n_{\mathbf{r},t}^{\text{cl},j}.
\end{aligned}
\label{eqn:Save}
\end{equation}
The first term ($iS_{\msf{ave1}}$), generated by disorder averaging, is nonlocal in time due to the static nature of the impurity potential. In contrast, the second term ($iS_{\msf{ave2}}$), originating from measurement-noise averaging, is local in time because $V_{\msf{meas}}$ is $\delta$-correlated in time.
Hubbard-Stratonovich decouplings are then applied separately to these two interaction terms using two auxiliary matrix fields: a bilocal-in-time matrix field $Q_{tt'}$ for $S_{\msf{ave1}}$ and a local-in-time matrix field $q_{t}$ for $S_{\msf{ave2}}$.
The different time structure of the two matrix fields is determined by that of the interactions they decouple.

For the disorder-averaging generated interaction $S_{\msf{ave1}}$, we decouple in the diffuson channel with the bilocal field $Q$: 
\begin{equation}
\begin{aligned}
e^{iS_{\msf{ave1}}}=\int\mathcal{D}Q\exp&\left\{-\frac{\pi \nu}{4\tau_{\text{el}}}\int_{t,t^\prime,\mathbf{r}}\Tr(\hat Q_{t t^\prime,\mathbf{r}}\hat Q_{t^\prime t,\mathbf{r}})\right.\\
&\left.-\frac{1}{2\tau_{\text{el}}}\int_{t,t^\prime,\mathbf{r}}{\bar\psi_{t,\mathbf{r}}\hat Q_{t t^\prime,\mathbf{r}}\psi_{t^\prime,\mathbf{r}}}\right\}.
\end{aligned}
\label{eqn:HS-dis}
\end{equation}
Note that we ignore the decoupling in the Cooperon channel, as we expect that the measurement breaks time-reversal symmetry and destroys the phase coherence between time-reversed paths. This measurement-induced dephasing effect appears as a mass term in the Cooperon, and suppresses the contribution of the Cooperon to the long-time physics.

The interaction $S_{\msf{ave2}}$ generated by measurement noise averaging
admits both an exchange decoupling channel and a direct decoupling channel \cite{Buchhold_free_2D_2024}.
Both channels contribute, and their relative weight $f$ is fixed by the causal structure of the saddle point (see Sec.~\ref{sec:sp}) and the $\mathrm{U}(1)$ Ward identity~\cite{Foster_interacting_2025}. This leads to
\begin{equation}
    \begin{aligned}
&e^{iS_{\msf{ave2}}}
=\int\mathcal{D}q
\exp\left\{
+i\int_{t,\mathbf{r}}{\bar\psi_{t,\mathbf{r}}\left[\hat{q}_{t,\mathbf{r}}-f\hat{\tau}^1\Tr{(\hat{q}_{t,\mathbf{r}}\hat{\tau}^1)}\right]\psi_{t,\mathbf{r}}}
\right.\\
&\left.
-\frac{1}{2g^2\alpha_f}\int_{t,\mathbf{r}}
\left[\Tr\left[(\hat{q}_{t,\mathbf{r}}\hat{\tau}^1)^2\right]-f\Tr^2{(\hat{q}_{t,\mathbf{r}}\hat{\tau}^1)}\right]
\right\},
\end{aligned}
\label{eqn:HS-meas}
\end{equation}
where  $\alpha_f=1/(f+1)$.

Combining both decouplings and integrating out the fermions, we arrive at an effective two-matrix theory:
\begin{equation}
\begin{aligned}
    Z[J]&=\int \mathcal{D}Q \mathcal{D}q \exp(iS),
    \\
    iS=&\Tr\ln\!\left[\hat{G}_0^{-1}+\hat{q}-f\hat{\tau}^1\Tr({\hat{q}\hat{\tau}^1})+\frac{i}{2\tel}\hat{Q}-\hat{\mathcal{J}}\right]\\
    &-\frac{1}{2g^2\alpha_{f}}\int_{t,\mathbf{r}}
    \left[\Tr\left[(\hat{q}_{t,\mathbf{r}}\hat{\tau}^1)^2\right]-f \Tr^2{(\hat{q}_{t,\mathbf{r}}\hat{\tau}^1)}\right]
    \\
    &-\frac{\pi \nu}{4\tau_{\text{el}}}\int_{t,t^\prime,\mathbf{r}}\Tr\left({\hat Q_{t^\prime t,\mathbf{r}}\hat Q_{t t^\prime,\mathbf{r}}}\right).
\end{aligned}
\label{eqn:SQq}
\end{equation}
Here ${Q}_{tt',\rb}^{ab,ij}$ 
is a  matrix in the Keldysh ($a,b$), replica ($i,j$), and time ($t,t'$) space, while ${q}^{ab,ij}_{t,\rb}$
is a  matrix in the Keldysh and replica spaces but is local in time.
The source matrix $\hat{\mathcal{J}}=\hat{\mathcal{J}}^\msf{cl}+\hat{\mathcal{J}}^\msf{q}\hat{\tau}^1$ is diagonal in the replica and time space, with  matrix elements $({\mathcal{J}}^{\msf{q/cl}})^{ij}_{tt'}(\rb)=J^{\msf{q/cl},i}(\rb,t)\delta_{ij}\delta_{tt'}$. We note that without disorder the $\hat{Q}$ field is absent, and the above equation reduces to the action of clean monitored fermions~\cite{Foster_interacting_2025}. Without the measurement the $\hat{q}$ field is absent, and it reduces to the standard theory of static disorder~\cite{Kamenev_2011,Liao_disorder_saddle_point_2017}.

\subsection{Saddle point}\label{sec:sp}

Taking the variation of the action in Eq.~\ref{eqn:SQq} with respect to $\hat{q}$ and $\hat{Q}$, separately, while setting the source field $J$ to zero, we obtain the saddle point equations:
\begin{subequations}\label{eqn:SPEQ}
    \begin{align}
        &\begin{aligned}
        &\frac{1}{g^2\alpha_f}\left[\hat{\tau}^1\hat q_{\msf{sp}}\hat{\tau}^1-f\hat{\tau}^1 \Trace{\hat{q}_{\msf{sp}}\hat{\tau}^1}\right]\lvert_{t,\mathbf{r}}
        \\
        &=\left[\hat G_{\msf{sp}}-f \hat{\tau}^1\Trace{\hat{G}_{\msf{sp}}\hat{\tau}^1}\right]\big\lvert_{(t,t;\mathbf{r},\mathbf{r})}, \label{eqn:SPEQ-q}
        \end{aligned}
        \\
    &\hat Q_{\msf{sp}}({t,t^\prime;\mathbf{r}})=\frac{i}{\pi \nu}\hat G_{\msf{sp}}(t,t^\prime;\mathbf{r},\mathbf{r}), \label{eqn:SPEQ-Q}
    \end{align}
\end{subequations}
where
\begin{equation}
    \hat{G}_{\msf{sp}}\equiv \left[G_0^{-1}+\hat{q}_{\msf{sp}}-f\hat{\tau}^1\Trace{\hat{q}_{\msf{sp}}\hat{\tau}^1}+\frac{i}{2\tel}\hat{Q}_{\msf{sp}}\right]^{-1}.
    \label{eqn:G}
\end{equation}

We use a saddle point ansatz in which both $\hat{Q}_\text{sp}$ and $\hat{q}_\text{sp}$ are diagonal in replica space, spatially uniform, and time translationally invariant. For simplicity, we focus on the half-filling case ($\int_{\e} F_0(\e)=0$), since the filling factor has been shown not to qualitatively affect the monitored dynamics~\cite{Mirlin_free_1D_2023,Foster_interacting_2025}.
The saddle points $\hat{Q}_\text{sp}$ and $\hat{q}_\text{sp}$ are then diagonal in Keldysh space and assume the form
\begin{equation}
    \hat{q}_\text{sp}(\omega,\mathbf{k})=i\gamma_{\msf{m}} \hat{\tau}^3\delta_{\ww,0}\delta_{\qb,0}, \quad \hat{Q}_\text{sp}(\e,\e';\mathbf{k})=\hat{\tau}^3 \delta_{\e,\e'}\delta_{\qb,0}.
    \label{eqn:SP}
\end{equation}
Here 
$\gamma_{\msf{m}}=\frac{g^2R}{2(1+R)}\left(\frac{\Lambda_k}{2\pi}\right)^d$ is the decay rate induced by measurement, with $\Lambda_k$ being the ultraviolet momentum cutoff. It is the measurement counterpart of the impurity-induced decay rate (elastic scattering rate) $\gamma_{\msf{D}}=1/2\tel$.

We note that the relative decoupling weight is set to $f=1/R$ by preserving the causal structure of the fermionic Green's function (in particular the vanishing of the ``anti-Keldysh" ($\msf{qq}$) component). This choice also ensures the $\mathrm{U}(1)$ Ward identity~\cite{Foster_interacting_2025} and reduces to a symmetric decoupling~\cite{Buchhold_free_2D_2024} in the limit $R\rightarrow 1$. Once $f$ is determined, the retarded and advanced ($\msf{clq}$ and $\msf{qcl}$) components of the saddle points follow from a self-consistent equation as in conventional disorder problem~\cite{Kamenev_2011}, while the Keldysh ($\msf{clcl}$) component is determined by the particle density at the saddle point and vanish at half filling~\cite{Foster_interacting_2025}.

This vanishing Keldysh components of both saddle points $\hat{Q}_{\msf{sp}}$ and $\hat{q}_{\msf{sp}}$ reflect the measurement-induced heating to an infinite temperature steady state~\cite{Mirlin_free_1D_2023,Mirlin_free_2D_2024,Foster_interacting_2025}. This is in contrast to the conventional disordered saddle point for $\hat{Q}$, whose Keldysh component is nonvanishing and set by the equilibrium fermionic Keldysh distribution function through the fluctuation-dissipation theorem~\cite{Kamenev_2011,Liao_disorder_saddle_point_2017}. Inserting the saddle point solutions into Eq.~\ref{eqn:G}, one can see that the disorder- and measurement-induced decay rates enter the saddle point Green's function $\hat{G}_{\msf{sp}}$ on equal footing, leading to a total decay rate  $\tilde{\gamma}=\gamma_{\msf{m}}+\gamma_{\msf{D}}$.

\section{Gaussian Approximation}\label{Sec:Gau}

To analyze the effective theory of monitored disordered fermions in Eq.~\ref{eqn:SQq}, we begin by examining the Gaussian fluctuations around the saddle points. In the limit of weak measurement and disorder strengths $\gamma_{\msf{m}},\gamma_{\msf{D}}\rightarrow 0$, the dominant contributions arise from the Gaussian fluctuations  $\delta \hat{q}=(\hat{q}-\hat{q}_{\msf{sp}})/(i\gamma_{\msf{m}})$ and $\delta \hat{Q}=\hat{Q}-\hat{Q}_{\msf{sp}}$, governed by the action
\begin{equation}
\begin{aligned}
    &iS^{(2)}
    =
    \frac{1}{2}\pi\nu\gamma_{\msf{m}}
    \int_{\e} 
    \int_{\omega,\mathbf{k}}
    \tr_R\left(
    \delta \bar{q}^{12}_{\omega,\mathbf{k}}
    \delta \bar{q}^{12}_{-\omega,-\mathbf{k}}
    +
    \delta \bar{q}^{21}_{\omega,\mathbf{k}}
    \delta \bar{q}^{21}_{-\omega,-\mathbf{k}}
    \right)
    \\
    &-\frac{\pi\nu}{2R}\gamma_{\msf{m}}
    \int_{\e} 
    \int_{\omega,\mathbf{k}}\!\!\!\!\!\!
    \tr_R\left(
    \delta \bar{q}^{12}_{\omega,\mathbf{k}}
    +
    \delta \bar{q}^{21}_{\omega,\mathbf{k}}
    \right)
    \tr_R\left(
    \delta \bar{q}^{12}_{-\omega,-\mathbf{k}}
    +
    \delta \bar{q}^{21}_{-\omega,-\mathbf{k}}
    \right)
    \\
    &-\pi\nu\gamma_{\msf{D}}
    \int_{\varepsilon,\omega,\mathbf{k}}
    \tr_R\left(
    \delta Q^{12}_{\e,\e-\ww;\mathbf{k}}
    \delta Q^{21}_{\e-\ww,\e;-\mathbf{k}}
    \right)
    \\
    &-\frac{\pi\nu}{\tilde{\gamma}}
    \int_{\varepsilon,\omega,\mathbf{k}}
    \!\!\!\!\!\!
    \left[-1+\theta(\omega,k)-\phi(\omega)\right]
    \\
    & \times
    \tr_{R}\left\lbrace
    \left[
    \gamma_{\msf{m}} \delta \bar{q}^{12}_{\omega,\mathbf{k}} 
    +
    \gamma_{\msf{D}} \delta Q^{12}_{\e,\e-\ww;\mathbf{k}}
    +
    i\mathcal{J}^{\msf{q}}_{\omega,\mathbf{k}}
    \right]
    \right.
    \\
    &\qquad \times
    \left.
    \left[
    \gamma_{\msf{m}}\delta  \bar{q}^{21}_{-\omega,-\mathbf{k}}
    +
    \gamma_{\msf{D}} \delta Q^{21}_{\varepsilon-\omega,\varepsilon;-\mathbf{k}}
    +
    i\mathcal{J}^{\msf{q}}_{-\omega,-\mathbf{k}}
    \right]\right\rbrace.
\end{aligned}
\label{eqn:S2}
\end{equation}
Here, for convenience, we have applied the transformation:
\begin{align}
    \hat{\bar{q}}=\hat{q}-f\hat{\tau}^1\Tr(\hat{q}\hat{\tau}^1),
    \qquad
    \delta \hat{\bar{q}}=(\hat{\bar{q}}-\hat{\bar{q}}_{\msf{sp}})/(i\gamma_{\msf{m}}).
\end{align} 
$\theta$ and $\phi$ are defined as 
\begin{equation}
    \theta(\omega,k)=\frac{{v}^2k^2}{4d\tilde\gamma^2}+\frac{\omega^2}{4\tilde\gamma^2}, \quad \phi(\omega)=\frac{i\omega}{2\tilde\gamma},
\end{equation}
where 
$v$ stands for the root-mean-square band velocity.

We focus on the fluctuation $\delta \hat{Q}^{12/21}$ and $\delta \hat{\bar{q}}^{12/21}$ that are massless in the disorder-only (disordered unmonitored) and measurement-only (clean monitored) theories. 
In the presence of both disorder and measurement, these modes acquire masses at the bare Gaussian level. However, once the coupling between the $\delta \hat{\bar{q}}$ and $\delta \hat{Q}$ is taken into account, a massless mode emerges corresponding to a particular linear combination of these two fields, as we show below.
By contrast, the Gaussian fluctuations $\delta \bar{q}^{11/22}$ and $\delta \hat{Q}^{11/22}$ are massive in the measurement-only and disorder-only cases, and they remain massive when both are present. These massive modes do not contribute to the long-time physics and are therefore not included in our analysis.



Note that the bilocal field $\delta \hat{Q}_{\e,\e-\ww}$ carries an extra frequency index relative to the local field $\delta \hat{\bar{q}}_{\ww}$. To handle this, we introduce a Fourier transform with respect to the center frequency $\e$, 
$
    \delta \hat{\mathcal{Q}}_{\ww,\mathbf{k}}(\Delta t)
    =
    \Lambda_{\e}^{-1}\int_{\e} \delta \hat{Q}_{\e,\e-\ww;\mathbf{k}} e^{-i\e \Delta t},
$
    with $\Lambda_\e=\int_\e=\frac{1}{2\pi\nu}\left(\frac{\Lambda_k}{2\pi}\right)^d$.
Here $\Delta t$ is the relative time in $\delta \hat{Q}_{t+\Delta t,t}$, so $\Delta t=0$ is the time-diagonal component.
at quadratic order, the nonlocal-in-time components with $\Delta t\neq 0$ are decoupled from the local-in-time component $\Delta t=0$ as well as $\delta \hat{\bar{q}}$ and $\hat{\mathcal{J}}$, because the latter are all local in time. We therefore ignore the $\Delta t\neq 0$ components in our Gaussian analysis.
Appendix~\ref{sec:app-dp-2} provides a complete treatment retaining all modes, including the nonlocal-in-time components of $\delta \hat{Q}$, which yields the same results.

We then decompose the fields
into replica-symmetric ($s$) and replica-asymmetric ($a$) components:
\begin{align}\label{eqn:para-perp}
    \hat{A}^{(s)}=\frac{\mathds{1}_R}{R} \tr_R \hat{A},
    \quad
    \hat{A}^{(a)}=\hat{A} -\hat{A}^{(s)},
    \quad
    \hat{A}=\delta \hat{\bar{q}} ,\delta \hat{Q}, \hat{\mathcal{J}}.
\end{align}
Here $\tr_R$ and $\mathds{1}_R$ denote the trace and identity matrix in the replica space, respectively.
The source field's replica-symmetric and replica-asymmetric components $J^{(s)}$ and $J^{(a)}$ couple separately to the particle density components in the corresponding sectors defined as
$
        n^{(s)}=\sum_{i}^R n^i/R$,
and $ 
        n^{(a)i}=n^{i}-n^{(s)}
$ (see $S_J$ in Eq.~\ref{eq:Z}).
$J^{(s/a)}$ is defined analogously and corresponds to the diagonal matrix element of the matrix $\hat{\mathcal{J}}^{(s/a)}$.
At the Gaussian level, the replica-symmetric and replica-asymmetric sectors decouple and can be analyzed independently.

The replica-symmetric sector determines the average-density response function, which describes the response of the average density to an external field,
\begin{align}\label{eq:Pi-para}
\begin{aligned}
    \Pi^{(s)ab}(\kb,\ww)=&
    \frac{\delta n^{a(s)}_{\ww,\kb}}{\delta J^{\bar{b}(s)}_{\ww,\kb}}\Bigg\lvert_{J=0}
    =
    -iR
    \braket{n^{a(s)}_{\ww,\kb}n^{b(s)}_{-\ww,-\kb}}
    \\
    =
    &
    \frac{i}{R}\frac{\delta^2 Z[J]}{\delta J^{\bar{a}(s)}_{-\ww,-\kb}\delta J^{\bar{b}(s)}_{\ww,\kb}}\Bigg\lvert_{J=0},
    \quad
    a,b=\msf{cl},\msf{q}.
\end{aligned}    
\end{align}
Here $\bar{a}$ denotes the complementary Keldysh index defined by $\bar{\msf{cl}}=\msf{q}$, $\bar{\msf{q}}=\msf{cl}$.

In contrast, the connected two-point density correlation function
\begin{align}\label{eq:twopoint}
\begin{aligned}
    &C(\rb,\rb';t,t')
    =
    \overline{
    \left\langle 
    \hat{n}_{\rb,t}\hat{n}_{\rb',t'}
    \right\rangle}
    -
    \overline{
    \left\langle 
    \hat{n}_{\rb,t}\right\rangle
    \left\langle \hat{n}_{\rb',t'}
    \right\rangle},
\end{aligned}
\end{align}
is determined by the replica-asymmetric sector.
Specifically, the first and second terms in the equation above correspond, respectively, to the replica-diagonal ($i=j$) and off-diagonal ($i\neq j$) components of the density correlation function
$\braket{n^{\msf{cl},i}_{\rb,t}n^{\msf{cl},j}_{\rb',t'}}$.
Combining these two parts shows that $C(\rb,\rb';t,t')$ reduces to the replica-asymmetric sector correlator and can be obtained from
\begin{align}\label{eq:Ca}
    \begin{aligned}    
    C(\rb,\rb';t,t')
    =&
    \frac{1}{R-1}
    \sum_{i}\left\langle
     n^{\msf{cl}(a),i}_{\rb,t} n^{\msf{cl}(a),i}_{\rb',t'}
    \right\rangle
    \\
    =&
     -\frac{1}{R-1}\sum_{i} \frac{\delta^2 Z[J]}{\delta J^{\msf{q}(a),i}_{\rb,t}\delta J^{\msf{q}(a),i}_{\rb',t'}}\bigg\lvert_{J=0}.
    \end{aligned}
\end{align}
Once $C(\rb,\rb';t,t')$ is known, the second cumulant of the particle number in the subsystem $A$ is determined by
$
    C_A^{(2)}(t_f)=\int_{\rb,\rb'\in A} C(\rb,\rb';t_f,t_f)
$
and the associated entanglement entropy is given by $S_{A}=\frac{\pi^2}{3} C_A^{(2)}$ within the Gaussian approximation.

\subsection{Replica-symmetric sector: Physical density response}

The Gaussian action for the replica-symmetric fluctuations $\delta \hat{\bar{q}}^{(s)}$ and $\delta \hat{\mathcal{Q}}^{(s)}(0)$ is given in Appendix~\ref{sec:app-dp} (Eq.~\ref{eqn:S-U1}), and it shows that the source field $J^{(s)}$ couples to these two fields only through the linear combination
\begin{align}
    \begin{aligned}
    \delta \hat{\mathcal{Q}}^{+(s)}_{\ww,\mathbf{k}}
    =
    \gm \delta \hat{\bar{q}}_{\ww,\mathbf{k}}^{(s)}
    +
    \gd \delta \hat{\mathcal{Q}}^{(s)}_{\ww,\mathbf{k}}(0).
    \end{aligned}
\end{align}
whose propagator is
\begin{align}
\begin{aligned}\label{eqn:dress-Prop-sym}
\braket{
    \delta \mathcal{Q}^{+(s)21}_{-\omega,-\mathbf{k}}
    \delta \mathcal{Q}^{+(s)12}_{\omega,\mathbf{k}}
    }
    =
    \frac{\tilde{\gamma}}{R\pi \nu \Lambda_{\e}} 
    \dfrac{1}{\theta(\omega,k)-\phi(\omega)}.
\end{aligned}
\end{align}
While $\delta \hat{\bar{q}}^{(s)}$ and $\delta \hat{\mathcal{Q}}^{(s)}(0)$ are separately massive (see Appendix~\ref{sec:app-dp} for the explicit expressions of their dressed propagators), their linear combination $\delta \hat{\mathcal{Q}}^{+(s)}$ is massless. Its propagator identical to that of $\gamma_{\msf{m}}\delta \hat{\bar{q}}^{(s)}$ in the measurement-only theory, except that the measurement-induced decay rate $\gm$ now replaced by the total decay rate $\tilde{\gamma}$.

Using Eq.~\ref{eq:Pi-para}, it is easy to see that the average-density response function $\Pi^{(s)ab}$ has only a single non-vanishing component---the Keldysh component $\Pi^{(s)K}\equiv\Pi^{(s)\msf{cl},\msf{cl}}$ determined by the propagator of $\delta \hat{\mathcal{Q}}^{+(s)}$.
In the low energy limit ($\ww,v k\ll \tilde{\gamma}$), $\Pi^{(s)K}(\ww,\kb)$ takes the usual diffusive form
\begin{equation}
\begin{aligned}
    i\Pi^{(s)K}(\ww,\kb)
    =&
     \dfrac{\frac{2\pi\nu \Lambda_{\e}}{\tilde{\gamma}}\theta(\omega,k)}{\theta^2(\omega,k)-\phi^2(\omega)}
     \simeq
     \left(\frac{\Lambda_k}{2\pi}\right)^d
     \dfrac{2Dk^2}{(Dk^2)^2+\ww^2},
\end{aligned}
\label{eqn:Pi-K}
\end{equation}
with the diffusion constant $D$ determined by the total decay rate $\tilde{\gamma}$, $D \equiv v^2/(2d\tilde{\gamma})$. 
This result has the same form as in the clean monitored theory, with disorder entering only through the replacement ${\gamma}_{\msf{m}} \to \tilde{\gamma}$ (equivalently, a modification of the diffusion constant $D$). It also confirms the Ward identity for charge conservation $\Pi^{(s)K}(\ww,\kb\rightarrow 0)=0$~\cite{Foster_interacting_2025}. 


 \subsection{Replica-asymmetric sector: Entanglement entropy}

In the replica-asymmetric sector,  we further decompose the fluctuations $\delta \hat{\bar{q}}^{(a)}\,(\delta \hat{\mathcal{Q}}^{(a)}(0))$ into its $\hat{\tau}^1$ and $\hat{\tau}^2$ components, denoted by $\hat{X}^{(\msf{m})}$ ($\hat{\mathcal{X}}^{(\msf{D})}$) and $\hat{Y}^{(\msf{m})}$ ($\hat{\mathcal{Y}}^{(\msf{D})}$), respectively. 
More specifically, we have
\begin{align}
\begin{aligned}
    &\hat{X}^{(\msf{m})}=(\delta \hat{\bar{q}}^{(a)12}+\delta \hat{\bar{q}}^{(a)21})/2,
    \\
    &\hat{Y}^{(\msf{m})}=i(\delta \hat{\bar{q}}^{(a)12}-\delta \hat{\bar{q}}^{(a)21})/2,
\end{aligned}
\end{align}
and $\hat{\mathcal{X}}^{(\msf{D})}$ and $\hat{\mathcal{Y}}^{(\msf{D})}$ are defined analogously in terms of $\delta \hat{\mathcal{Q}}^{(a)}(0)$.
These fluctuations are governed by the action in Eq.~\ref{eqn:S-SUR} in Appendix~\ref{sec:app-dp}.

In the absence of disorder, $\hat{Y}^{(\msf{m})}$ is massless while $\hat{X}^{(\msf{m})}$ is massive, and integrating out the massive $\hat{X}^{(\msf{m})}$ mode only leads to a renormalization of the parameters. By contrast, in the disorder-only case, both $\hat{\mathcal{X}}^{(\msf{D})}$ and $\hat{\mathcal{Y}}^{(\msf{D})}$ are massless.
In the current case, when both disorder and measurement are present, all four modes are massive. However, their coupling results in a massless mode corresponding to a linear combination of $\hat{Y}^{(\msf{m})}$ and $\hat{\mathcal{Y}}^{(\msf{D})}$, while the sector spanned by $\hat{X}^{(\msf{m})}$ and $\hat{\mathcal{X}}^{(\msf{D})}$  remains massive, as in the measurement-only case.

As in the replica-symmetric sector, we introduce
\begin{align}
\begin{aligned}
    \hat{Y}^{(+)}_{\omega,\mathbf{k}}
    =
    \gamma_{\msf{m}} \hat{Y}^{(\msf{m})}_{\omega,\mathbf{k}} 
    +
    \gamma_{\msf{D}}  \hat{\mathcal{Y}}^{(\msf{D})}_{\ww,\mathbf{k}},
  \end{aligned}  
\end{align}
and similarly for $\hat{X}^{(+)}$. These modes  couple directly to the source field $J^{(a)}$.
We find that the propagator for $\hat{Y}^{(+)}$ is massless while that for $\hat{X}^{(+)}$ is massive: 
\begin{align}\label{eqn:dres-Prop}
\begin{aligned}
    &\braket{
    Y_{-\omega,-\mathbf{k}}^{(+)ij}
    Y_{\omega,\mathbf{k}}^{(+)uv}
    }
    =
    \frac{P_{ij,uv}}{2 \pi \nu \Lambda_{\e}} 
    \frac{\tilde{\gamma}}{\theta(\omega,k)},
    \\
    &\braket{
    X_{-\omega,-\mathbf{k}}^{(+)ij}
    X_{\omega,\mathbf{k}}^{(+)uv}
    }
     =
   \frac{P_{ij,uv}}{2 \pi \nu \Lambda_{\e}} 
   \frac{\gd^2-\gm^2}{2\gm},
\end{aligned}
\end{align}
where $P_{ij,uv} \equiv \left(\delta_{iv}\delta_{ju}-\frac{1}{R}\delta_{ij}\delta_{uv}\right)$.

Integrating out the massive mode $\hat{X}^{(+)}$, we arrive at an effective Gaussian action for massless mode $\hat{Y}^{(+)}$:
 \begin{align}
 \begin{aligned}\label{eq:S_z}
    S^{(2a)\msf{eff}}
    =&
    -\frac{\pi\nu}{\tilde{\gamma}}
    \Lambda_{\e}
    \int_{\omega,\mathbf{k}}
     \tilde{\theta}(\ww,k)
    \tr_R\left(
    \hat{Y}^{(+)}_{-\omega,-\mathbf{k}}
    \hat{Y}^{(+)}_{\omega,\mathbf{k}}
   \right)
    \\
    &+
    z\frac{2\pi\nu}{\tilde{\gamma}}
    \Lambda_{\e}
    \int_{\omega,\mathbf{k}}
     \phi(\ww)
    \tr_R\left(
    \hat{\mathcal{J}}^{\msf{q}(a)}_{-\omega,-\mathbf{k}}
    \hat{Y}^{(+)}_{\omega,\mathbf{k}}
    \right)
     \\
    &-z
    \frac{\pi\nu}{\tilde{\gamma}}
    \Lambda_{\e}
    \int_{\omega,\mathbf{k}}
    \tr_R\left(
    \hat{\mathcal{J}}^{\msf{q}(a)}_{-\omega,-\mathbf{k}}
    \hat{\mathcal{J}}^{\msf{q}(a)}_{\omega,\mathbf{k}}
    \right),
\end{aligned}
\end{align}
where $z= \tilde{\gamma}/{2\gm}$ accounts for the massive mode renormalization effect and $\tilde{\theta}(\ww,k)$ is defined as
\begin{align}
\begin{aligned}
    \tilde{\theta}(\ww,k)
    =
    \frac{{v}^2k^2}{4d\tilde\gamma^2}+z\frac{\omega^2}{4\tilde\gamma^2}.
\end{aligned}
\end{align}
The renormalized $\hat{Y}^{(+)}$ propagator  retains the same form in Eq.~\ref{eqn:dres-Prop} with $\theta(\ww,k)$ replaced by $\tilde{\theta}(\ww,k)$.

Using Eq.~\ref{eq:S_z} and Eq.~\ref{eq:Ca},  the connected density correlation function $C(\rb,\rb';t,t')$ is determined  by the $\hat{Y}^{(+)}$ propagator
\begin{align}
    \begin{aligned}    
    &C(\rb,\rb';t,t')
     =
    z\frac{2\pi\nu\Lambda_{\e}}{\tilde{\gamma}}
    \delta(\rb-\rb')\delta(t-t')
    \\
    &
    +z^2
    (\frac{2\pi\nu\Lambda_{\e}}{\tilde{\gamma}})^2
    \int_{0}^{\infty}  \frac{d\ww}{\pi}
    \cos\left(\ww (t-t_f)\right)\cos\left(\ww (t'-t_f)\right)
    \\
    &\qquad\qquad \times
    \int_{\kb} e^{i\kb(\rb-\rb')}
    \left(
    \frac{\phi^2(\ww)}{R-1} 
    \sum_{i} 
    \braket{
     Y_{-\kb,-\ww}^{(+)ii}
     Y_{-\kb,-\ww}^{(+)ii}
    }\right),
        \end{aligned}
\end{align}
Here the factor $\cos\left(\ww (t-t_f)\right)$ originates from the half-space Fourier transform of $\partial_t \hat{Y}_t^{(+)}$
with the boundary condition $\hat{Y}^{(+)}_{t_f}=0$ at the final time $t_f=0$~\cite{Mirlin_ineracting_2025}.
Substituting the explicit expression of the renormalized $Y$ propagator and setting $t=t'=t_f$, we obtain the equal-time correlator $C(\rb,\rb';t_f,t_f)$, whose Fourier transform is
\begin{align}
    \begin{aligned} 
    &\tilde{C}(\kb)
    =\frac{2\nu \Lambda_{\e}}{\tilde{\gamma}}
    \int_{0}^{\infty}  d\ww
    \left[
    z
    +
    z^2\frac{ \phi^2(\ww)}{\tilde{\theta}(\omega,k)}
    \right]
    =
    \left(\frac{\Lambda_k}{2\pi}\right)^d
    \frac{v}{\tilde{\gamma}}
    \sqrt{\frac{z}{d}}k.
    \end{aligned}
\end{align}
For a subsystem of linear size $\ell$, this leads to the entanglement entropy
\begin{align}
    \begin{aligned}
        S_A \propto   \left(\frac{\Lambda_k}{2\pi}\right)^d \frac{v}{\tilde{\gamma}}\sqrt{\frac{z}{d}}\ell^{d-1}\ln (\ell/\ell_0),
    \end{aligned}
    \label{eq:SA}
\end{align}
up to a geometry-dependent overall coefficient, with the effective mean free path $\ell_0 \simeq v/\tilde{\gamma}$ serves as the ultraviolet cutoff. 

As in the case of the physical response function, the entanglement entropy retains the same form as the clean monitored theory, with disorder entering only through the replacements $\gamma_{\msf{m}} \to \tilde{\gamma}$ and $z=1/2 \to z=\tilde{\gamma}/2\gamma_{\msf{m}}$. Note that in the regime $\gamma_{\msf{m}}/\gamma_{\msf{D}} \ll 1$, the neglected nonlocal-in-time modes of $\delta \hat{Q}$ (see Appendix~\ref{sec:app-dp-2}) become relevant, and a separate analysis is required, as discussed in Sec.~\ref{sec:NLSM}.

\section{Nonlinear Sigma model}\label{sec:NLSM}

The two-matrix field  theory in Eq.~\ref{eqn:SQq} is governed by $\hat{q}_{t,\rb}$ and $\hat{Q}_{tt',\rb}$, introduced to decouple separately interactions generated by disorder and measurement averaging, and carrying different structures in time space. However, the symmetry of the action in Eq.~\ref{eqn:SQq} implies that the Goldstone manifold of the theory is generated by simultaneous unitary rotations of the matrix fields $\hat{Q}$ and $\hat{\bar{q}}$
around their respective saddle points, both of which are local in time, by the same local-in-time unitary matrix  $\hat{\mathcal{R}}_{t,\rb}$ 
 \begin{align}\label{eq:Goldstone}
 \begin{aligned}
\hat{Q}_{tt',\rb}=\hat{\mathcal{R}}_{t,\rb}\hat{\tau}^3\hat{\mathcal{R}}_{t,\rb}^{-1}\delta_{tt'},
     \qquad
     \hat{\bar{q}}_{t,\rb}=i\gm \hat{{\mathcal{R}}}_{t,\rb}\hat{\tau}^3\hat{\mathcal{R}}_{t,\rb}^{-1}.
 \end{aligned}
 \end{align}
The rotation matrix $\hat{\mathcal{R}}_{t,\rb}$ can be parameterized as
\begin{align}\label{eq:R}
\begin{aligned}
    & \hat{\mathcal{R}}=\hat{\mathcal{R}}_{\Theta}\hat{\mathcal{R}}_0, 
    \quad
    \hat{\mathcal{R}}_0=\hat{\mathcal{R}}_{\theta}\hat{\mathcal{R}}_{\phi}
     \\
    &\hat{\mathcal{R}}_{\Theta}=e^{i\hat{\Theta} \otimes \hat{\tau}^1/2},
    \quad
    \hat{\mathcal{R}}_{\theta}=e^{i\theta \hat{\tau}^1/2},
    \quad
    \hat{\mathcal{R}}_{\phi}=e^{i\phi \hat{\tau}^2/2},
 \end{aligned}   
\end{align}
where $\hat{\Theta}$ is a traceless matrix field in replica space satisfying $\tr_R(\hat{\Theta})=0$. 
The matrix field $\hat{\Theta}$ generates rotations $\hat{\mathcal{R}}_{\Theta}$ in the replica-asymmetric sector, while the scalar fields $\theta$ and $\phi$ generate rotations $\hat{\mathcal{R}}_{0}$ in the replica-symmetric sector.
As in the measurement-only case, the rotation $\hat{\mathcal{R}}_{\phi}$ produces massless fluctuations due to the specific choice of the decoupling weight $f=1/R$. By contrast, in the replica-asymmetric sector, an analogous rotation $\hat{\mathcal{R}}_{\Phi}=e^{i \hat{\Phi} \otimes \hat{\tau}^2/2}$, with $\tr_R(\hat{\Phi})=0$,  generates massive modes when applied to both $\hat{Q}$ and $\hat{\bar{q}}$ matrix fields. 

The structure of this Goldstone manifold can be understood by comparing the disorder-only and measurement-only theories. The disorder-only theory involves only $\hat{Q}$, and its action is invariant under unitary rotations of  $\hat{Q}$ in replica, Keldysh, and time space (ignoring the cost from $\hat{G}_0^{-1}$, which enters only through gradients and time derivatives), including nonlocal-in-time rotations that mix different time indices. The fluctuations generated by such unitary rotations that do not leave the saddle point invariant are therefore Goldstone modes~\cite{Kamenev_2011}. For the measurement-only theory, which involves only the local-in-time matrix field $\hat{q}$, the rotations that preserve the actions are local in time and are further constrained by the $\hat{\tau}^1$ structure in the quadratic action. The Goldstone manifold is generated by local-in-time unitary rotations parameterized by Eq.~\ref{eq:R}~\cite{Mirlin_free_1D_2023,Mirlin_free_2D_2024,Buchhold_free_2D_2024}. When both the measurement and disorder are present, the action remains invariant only when both fields are rotated by the same local-in-time rotation. The surviving symmetry,  and therefore the Goldstone manifold, coincide with that of the measurement-only theory.



The soft modes of the measurement-only or disorder-only theories that lie outside this Goldstone manifold now acquire a mass when both disorder and measurement are present, and they can be divided into three different categories.
The first type arises when both $\hat{Q}$ and $\hat{q}$ are rotated by the same unitary matrix $\hat{\mathcal{R}}_{\Phi}=e^{i\Phi\otimes \hat{\tau}^2/2}$. The action is then no longer invariant, because of the $\hat{\tau}^1$ structure of the quadratic action for $\hat{q}$. As a result, fluctuations generated by such a rotation become massive.
The second type is generated by relative local-in-time rotations between the two fields. 
Acting on $\hat{Q}$ with an additional local-in-time unitary rotation $\tilde{\mathcal{R}}_{t}$, 
\begin{align}\label{eq:rr}
    \hat{Q}_{tt';\rb}=\hat{\mathcal{R}}_{t,\rb}\hat{\tilde{\mathcal{R}}}_{t,\rb}\hat{\tau}^3\hat{\tilde{\mathcal{R}}}^{-1}_{t,\rb}\hat{\mathcal{R}}^{-1}_{t,\rb}\delta_{tt'},
\end{align}
while keeping $\hat{\bar{q}}$ fixed as in Eq.~\ref{eq:Goldstone},  breaks the symmetry of the present theory and  produces massive modes. 
These first two types of massive modes couple to the Goldstone modes at quadratic order and renormalize the effective Goldstone action once integrated out.
The third type of massive modes arises from nonlocal-in-time rotations  $\hat{\mathcal{R}}^{Q}_{t''t',\rb}$ around the saddle point $\hat{Q}_{\msf{sp}}$, which mix different time indices and produce off-diagonal components in time space,
\begin{align}
    \hat{Q}_{tt',\rb}=\int_{t''}
    \hat{\mathcal{R}}^{Q}_{tt'',\rb}
    \hat{\tau}^3 
    \hat{\mathcal{R}}^{Q-1}_{t''t',\rb}.
\end{align}
The presence of the measurement-induced matrix field $\hat{\bar{q}}$ generates a finite mass for the off-diagonal-in-time components  $\delta \hat{Q}_{tt'}$ ($t\neq t'$), which are not included in either Eq.~\ref{eq:rr} or Eq.~\ref{eq:Goldstone}. Unlike the first two types, these nonlocal-in-time fluctuations decouple from the Goldstone sector at quadratic order, and therefore can be ignored in the regime of weak measurement and disorder strengths, except when $\gamma_{\msf{m}}/\gamma_{\msf{D}} \ll 1$ (see the discussion at the end of this section).



Let us first focus  on the Goldstone modes, temporarily neglecting the massive modes and the source field. The Goldstone action can be obtained by substituting the parameterization Eq.~\ref{eq:Goldstone} into Eq.~\ref{eqn:SQq},
\begin{equation}\label{eq:SGM-1}
\begin{aligned}
    iS_{gm}
    =&\Tr\ln\left[\hat{G}_0^{-1}+i\tilde{\gamma} \hat{\mathcal{R}} \hat{\tau}^3 \hat{\mathcal{R}}^{-1}\right].
\end{aligned}
\end{equation}
This expression immediately shows that the effective theory for the Goldstone fluctuations in the presence of both continuous monitoring and static disorder is identical to that of
the measurement-only theory, under the replacement $\gm \rightarrow \tilde{\gamma}$. The gradient expansion then proceeds exactly as in the measurement-only theory, yielding the same Goldstone action under this replacement (see Appendix~\ref{sec:app-nlsm-gm} for the detailed derivation).

To complete the derivation, we  then include the massive modes generated by the rotation $\hat{\mathcal{R}}_{\Phi}=e^{i\Phi\otimes \hat{\tau}^2/2}$, which acts on both $\hat{\bar{q}}$ and $\hat{Q}$, as well as those arising from the relative rotation $\hat{\tilde{\mathcal{R}}}$ between them (Appendix~\ref{sec:app-nlsm-mm}). Once the coupling to the source field $\hat{J}$ is included,  integrating out the massive modes renormalizes not only the Goldstone action, but also the coupling between the Goldstone modes and the source field, as well as the source-field action (Appendix~\ref{sec:app-nlsm-s}).

Collecting all contributions, we arrive at the effective action for the Goldstone modes $\hat{U}=\exp(i\hat{\Theta})\in \mathrm{SU}(R)$ in the replica-asymmetric  sector and $\hat{\mathcal{Q}}_0=\hat{\mathcal{R}}_0\hat{\tau}^3 \hat{\mathcal{R}}_0^{-1} \in \mathrm{U}(2)/\mathrm{U}(1)\times \mathrm{U}(1)$ in the replica-symmetric  sector: 
\begin{equation}\label{eq:SGM-F2}
\begin{aligned}
    &iS_{gm}'=
     \frac{R}{2}\left\lbrace
     \tr_K\left(\hat{\tau}^3 \hat{\mathcal{R}}^{-1}_0 \partial_t \hat{\mathcal{R}}_0\right)
    -\frac{ D}{4}
     \tr_K\left[(\nabla \hat{\mathcal{Q}}_0)^2\right]
     \right\rbrace
     \\
     &
     +\frac{iR}{2}
    \tr_{K}\left[\hat{\mathcal{J}}^{\msf{q}(s)}\hat{\tau}^1\hat{\mathcal{Q}}_0\right]
    -
    \frac{R}{2\tilde{\gamma}}\left[\hat{\mathcal{J}}^{\msf{q}(s)}\hat{\mathcal{J}}^{q(s)}\right]
    \\
    &-\frac{1}{8\tilde{\gamma}}\left\lbrace
    z
    \tr_{R} \left[ \left(\partial_t^J \hat{U}\right) \left(\partial_t^J \hat{U}\right)^{\dagger}\right]
    +
    \frac{v^2}{d}\tr_{R}\left(\nabla \hat{U}\nabla \hat{U}^{\dagger} \right)
    \right\rbrace,
\end{aligned}
\end{equation}
where
\begin{align}
\begin{aligned}
&\partial_t^J \hat{U}= \partial_t \hat{U}+i \left(\hat{\mathcal{J}}^{\msf{q}(a)}\hat{U}+\hat{U}\hat{\mathcal{J}}^{\msf{q}(a)}\right),
 \\
& \left(\partial_t^J U\right)^{\dagger}= \partial_t \hat{U}^{\dagger}-i \left(\hat{\mathcal{J}}^{\msf{q}(a)}\hat{U}^{\dagger}+\hat{U}^{\dagger}\hat{\mathcal{J}}^{\msf{q}(a)}\right).
 \end{aligned}
\end{align}
Here we have omitted an overall prefactor of $\left(\frac{\Lambda_k}{2\pi}\right)^d$ in the renormalized action $S_{gm}'$ for simplicity. 
$\tr_K$ denotes the trace over the Keldysh space only.
At quadratic order, this action agrees with the effective Gaussian actions $S^{(2a)\msf{eff}}$ (Eq.~\ref{eq:S_z}) and  $S^{(2s)}$ (Eq.~\ref{eqn:S-U1}) for the mixed modes  $\hat{Y}^{(+)}$ and $\delta\hat{\mathcal{Q}}^{(+)}$. 

We note that the derivation of the NL$\sigma$M for disordered monitored fermions presented here is controlled in the regime of sufficiently weak disorder and measurement strengths $\gamma_{\msf{m}},\gamma_{\msf{D}}\ll mv^2$. However, even when both $\gm$ and $\gd$ are small, in the limit $\gm/\gd \ll 1$, the neglected nonlocal-in-time massive modes $\delta \hat{Q}$ might become relevant. These modes are generated by nonlocal-in-time rotations around $\hat{Q}_{\msf{sp}}$. They are massless in the disorder-only limit and acquire a mass in the presence of measurement (see Appendix~\ref{sec:app-dp-2} for their Gaussian propagators). For $\gm/\gd \ll 1$, this mass becomes small so that these modes cannot be safely ignored despite being decoupled from the local-in-time Goldstone fluctuations at quadratic order. 
A more detailed analysis of the contribution of these massive modes in this regime is left for a future study.

The factor $z=\tilde{\gamma}/2{\gamma}_m$ in the action above arises from integrating out massive modes and reduces to $z=1$ if these modes are neglected. 
It modifies the relative coefficients of the temporal and spatial gradient terms and can be absorbed by a rescaling of time. 
Apart from this $z$ factor, which takes the value $z=1/2$ in the measurement-only theory, the resulting NL$\sigma$M for monitored disordered fermions is identical to that of the measurement-only theory, upon the replacement ${\gamma}_m \to \tilde{\gamma}$ (with the diffusion constant $D \propto 1/\tilde{\gamma}$ modified accordingly). 
The same sigma model also appears in conventional equilibrium disordered systems in chiral unitary class AIII in spatial dimension $d+1$~\cite{Anderson_Transition_Review}.

This result implies that the same renormalization group analysis performed for the measurement-only theory applies here as well, leading to the same one-loop RG equation~\cite{Mirlin_free_2D_2024,Buchhold_free_2D_2024}.
Since $z$ only modifies the relative coefficient of the temporal and spatial gradient terms, it can be absorbed by a rescaling of time and does not affect the universal RG flow. We therefore set $z=1$ for the following qualitative RG analysis.
The resulting stiffness of the present NL$\sigma$M is 
$\lambda=\frac{v}{4\sqrt{d}\tilde{\gamma}}\left(\frac{\Lambda_k}{2\pi}\right)^d$, and is determined by the inverse of the total decay rate $\tilde{\gamma}$ rather than the measurement-induced decay rate $\gm$ in the measurement-only theory. The corresponding dimensionless coupling constant $\tilde{\lambda}(l)=\lambda(l)l^{d-1}$ then obeys the RG equation for the AIII NL$\sigma$M~\cite{Anderson_Transition_Review,Mirlin_free_2D_2024,Buchhold_free_2D_2024}
\begin{align}
    \frac{d\tilde{\lambda}}{d\ln l}=(d-1) \tilde{\lambda}-\frac{R}{4\pi}
    +O(\tilde{\lambda}^{-1}).
\end{align}
Taking the replica limit $R\rightarrow 1$, this suggests the presence (absence) of an MIPT in dimension $d>1$ ($d=1$).

For dimension $d>1$, the RG equation has an unstable fixed point at $\tilde{\lambda}_c = 1/4\pi(d-1)$. The transition therefore occurs when 
$\tilde{\lambda}(\ell_0)=\tilde{\lambda}_c$, where $\ell_0 \simeq v/\tilde{\gamma}$ is the mean free path.
This condition gives a critical total decay rate $\tilde{\gamma}_c \sim (\frac{d-1}{\sqrt{d}})^{1/d} v\Lambda_k$, which is also the critical measurement rate in the clean case.
 Therefore, the critical measurement strength in the presence of disorder $\gm^c$ is related to the critical value in the clean case $\tilde{\gamma}_c$ by
 \begin{equation}\label{eq:gmc}
    \gm^c = \tilde{\gamma}_c
     - \gd\,.
\end{equation}
When the factor $z$ is retained, the bare stiffness after time rescaling is instead $\lambda=\sqrt{z}\frac{v}{4\sqrt{d}\tilde{\gamma}}\left(\frac{\Lambda_k}{2\pi}\right)^d$ and the linear relation above should be modified. Nevertheless, since $\lambda \propto \sqrt{z}/\tilde{\gamma}=1/\sqrt{2\gm(\gm+\gd)}$, increasing either $\gm$ and $\gd$ reduces the bare stiffness $\lambda$. Therefore, retaining $z$ does not change the qualitative prediction that disorder shifts the transition to a lower critical measurement strength.
In other words, we expect the disorder potential to enlarge the area-law regime, causing the entanglement transition to occur at weaker measurement rates in the presence of disorder.
As in the Anderson localization transition, this RG analysis is controlled for $(d-1) \ll 1$ and is expected to describe the transition qualitatively in dimensions $d=2,3$~\cite{Mirlin_free_2D_2024}.

In one dimension, the stiffness $\lambda$ decreases under RG and flows to the strong-coupling regime when the system size exceeds the correlation length,
\begin{equation}
    \ell_\mathsf{corr} \sim \ell_0 \exp{(4\pi\lambda_0)}\,,
\end{equation}
where $\lambda_0=\frac{v}{4\tilde{\gamma}}\left(\frac{\Lambda_k}{2\pi}\right)$ is the bare stiffness.
At length scales beyond this correlation length $\ell \gtrsim  \ell_{\mathsf{corr}}$, the logarithmic growth given in Eq.~\ref{eq:SA} is cut off as the RG flow enters the strong coupling regime, and the entanglement entropy saturates, indicating that the system is in the area-law phase~\cite{Mirlin_free_1D_2023}. The saturation value of the entanglement entropy is approximately,
\begin{equation}
    S_A (l \gtrsim l_\mathsf{corr}) \sim \left(\frac{\Lambda_k}{2\pi}\right)^{2} \frac{v^2}{\tilde{\gamma}^2} \,.
    \label{eq:SA_sat}
\end{equation}

 \section{Numerical Analysis}\label{sec:num}

\begin{figure}
    \centering
    \includegraphics[width=0.95\linewidth]{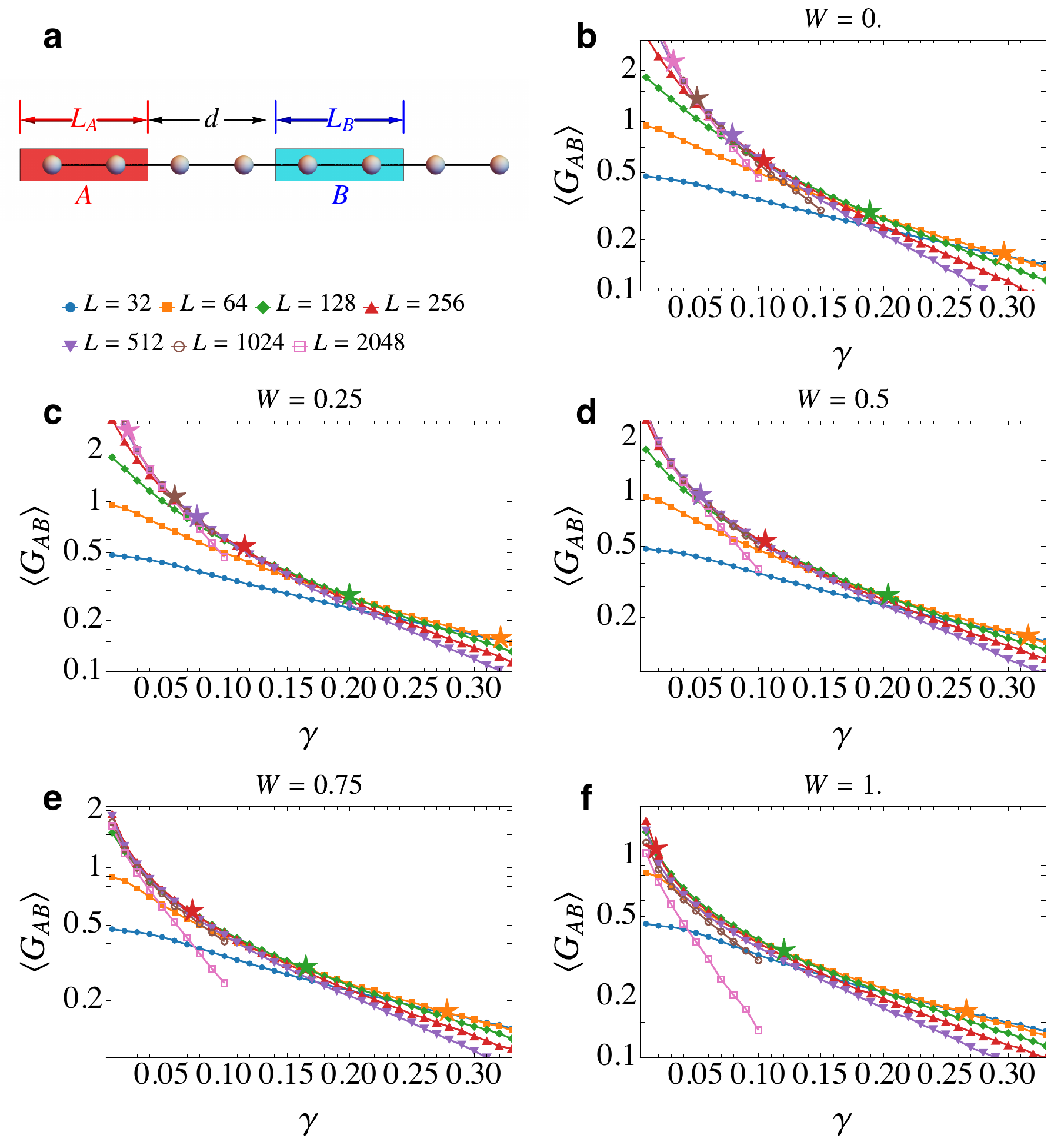}
       \label{fig:GAB-W}
       \caption{The steady-state covariance $G_{AB}$ for 1D disordered monitored fermions as a function of measurement rate for various disorder strength and system size. The geometry of $A$ and $B$ region in the 1D chain is illustrated in (a) and we have $L_A=L_B=d=L/4$. The stars in (b)--(f) represent the crossing point of adjacent system sizes and the color is the same as that of the larger one. The stars for  larger system sizes with strong disorder are not shown as $G_{AB}$ already show area-law decreasing with $L$ and no crossings are observed. The data is averaged over $1000$ quantum trajectories for $L=32,\cdots, 256$; $400$ quantum trajectories for $L=512$; $100$ quantum trajectories for $L=1024$; and $50$ quantum trajectories for $L=2048$.}
\end{figure}

To verify the analytical prediction, we perform numerical analysis on monitored dirty fermions. We consider disordered fermion on a 1D chain and a 2D square lattice, governed by the Hamiltonian,
  \begin{equation}
     H =   -t\sum_{\langle i,j\rangle}(c^\dagger_{i} c_{j} + H.c.) + \sum_{i} V_i n_i\,.
     \label{eq:H}
 \end{equation}
 Here $V_i$ is the onsite scalar potential, uniformly distributed in $[-W, W]$. We consider the quantum projection dynamics starting from an initial half-filled N\'{e}el state $\ket{\Psi_0} = \ket{1010\cdots 10}$. The fermions are subject to unitary evolution governed by the Hamiltonian in Eq.~\eqref{eq:H}. At every time step $\delta t$, a randomly chosen site $i$ is projected into the occupied state $\ket{1}$ with probability $\langle n_i\rangle$, or into the empty state $\ket{0}$ otherwise. Since the quantum state remains Gaussian throughout the dynamics, we are able to perform large-scale numerical simulations. The quantum projection dynamics is simulated until steady state is reached, at which the physical observables are measured and averaged over the steady states of many quantum trajectories. In this work, we employ a uniform measurement interval with $\delta t=1/(N\gamma)$ with $N$ being the number of the sites in the system. Here $\gamma$ denotes the measurement rate that is proportional to the measurement-induced decay rate $\gm$.

Although the quantum projection protocol differs from the continuous monitoring discussed in previous sections, both belong to the same symmetry class and are governed by the identical NL$\sigma$M as the one derived in Sec. \ref{sec:NLSM}. Therefore, the underlying physics of continuous and discrete monitoring is equivalent, allowing us to use either protocol to investigate  measurement-induced dynamics and potential phase transitions. We employ the quantum projection protocol in this work as it is computational more efficient and exhibits weaker finite-size effect than quantum state diffusion \cite{Cao_free_1D_numerical_2019, Diehl_free_1D_numerical_2021}.
The computational advantage is achieved by utilizing a rank-$1$ polar update algorithm for the re-orthogonalization. The rank-$1$ polar decomposition is mathematically equivalent to singular value decomposition (SVD), which best preserves the structure of $\ket{\Psi}$ during the re-orthogonalization. This rank-$1$ method is significantly faster than SVD or standard QR decomposition, and the Newton-Schulz QR is applied periodically to mitigate the accumulation of rounding errors. Further details are provided in Appendix~\ref{sec:app-numerics}.

The steady-state particle-number covariance $G_{AB}$, which is  proportional to the mutual information $I_{AB}\approx \frac{3\pi^2}{3}G_{AB}$ \cite{Mirlin_free_2D_2024}, is used to characterize possible MIPT in the quantum projection dynamics. The covariance $G_{AB}$ between two regions $A$ and $B$ is defined as
\begin{equation}
G_{AB} = - \int_{\vex{x}\in A} \int_{\vex{y}\in B} C(\vex{x}-\vex{y})\,.
\label{eq:GAB}
\end{equation}
Here the equal-time two-point correlation function [Eq.~\ref{eq:twopoint}] is given by,
\begin{equation}
    C(\vex{x}, \vex{y}) = \left<n(\vex{x}) n(\vex{y})\right> - \left<n(\vex{x})\right> \left<n(\vex{y})\right>\,.
    \label{eq:Cxy}
\end{equation}
The covariance is then averaged over quantum trajectories, disorder realizations, position, and  a time interval after steady condition is reached.

 \subsection{Monitored Dirty Fermions in 1D: Absence of MIPT}
 
\begin{figure}
    \centering
    \includegraphics[width=0.95\linewidth]{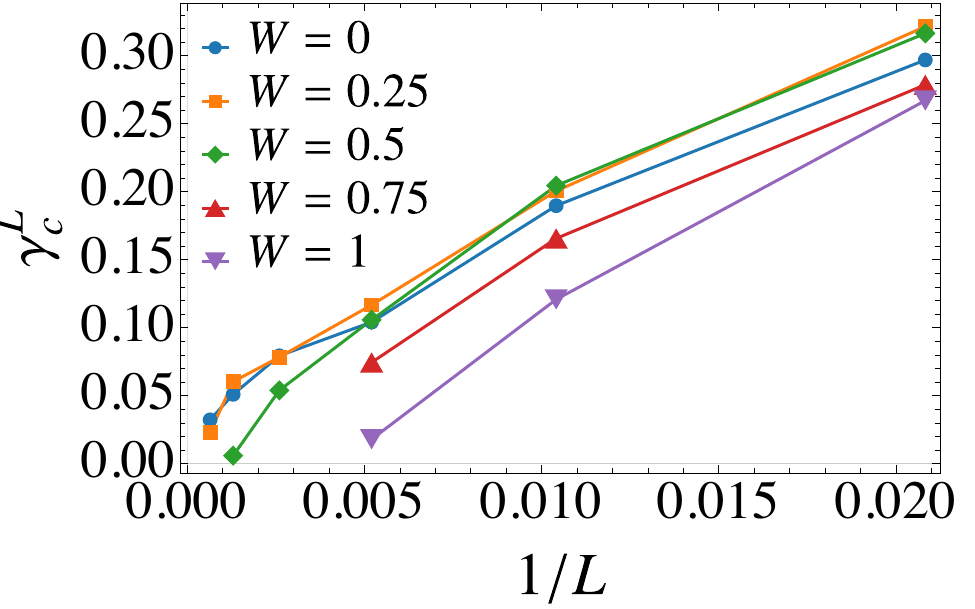}
    \caption{Finite size crossing point of $G_{AB}$ curves of adjacent system size in the 1D disordered monitored fermion system shown in Fig.~\ref{fig:GAB-W}. The decreasing of the crossing point  with increasing system size $L$ indicates the absence of genuine MIPT in thermodynamic limit.}
    \label{fig:gamma_c_l}
\end{figure}

\begin{figure}
    \centering
    \includegraphics[width=0.95\linewidth]{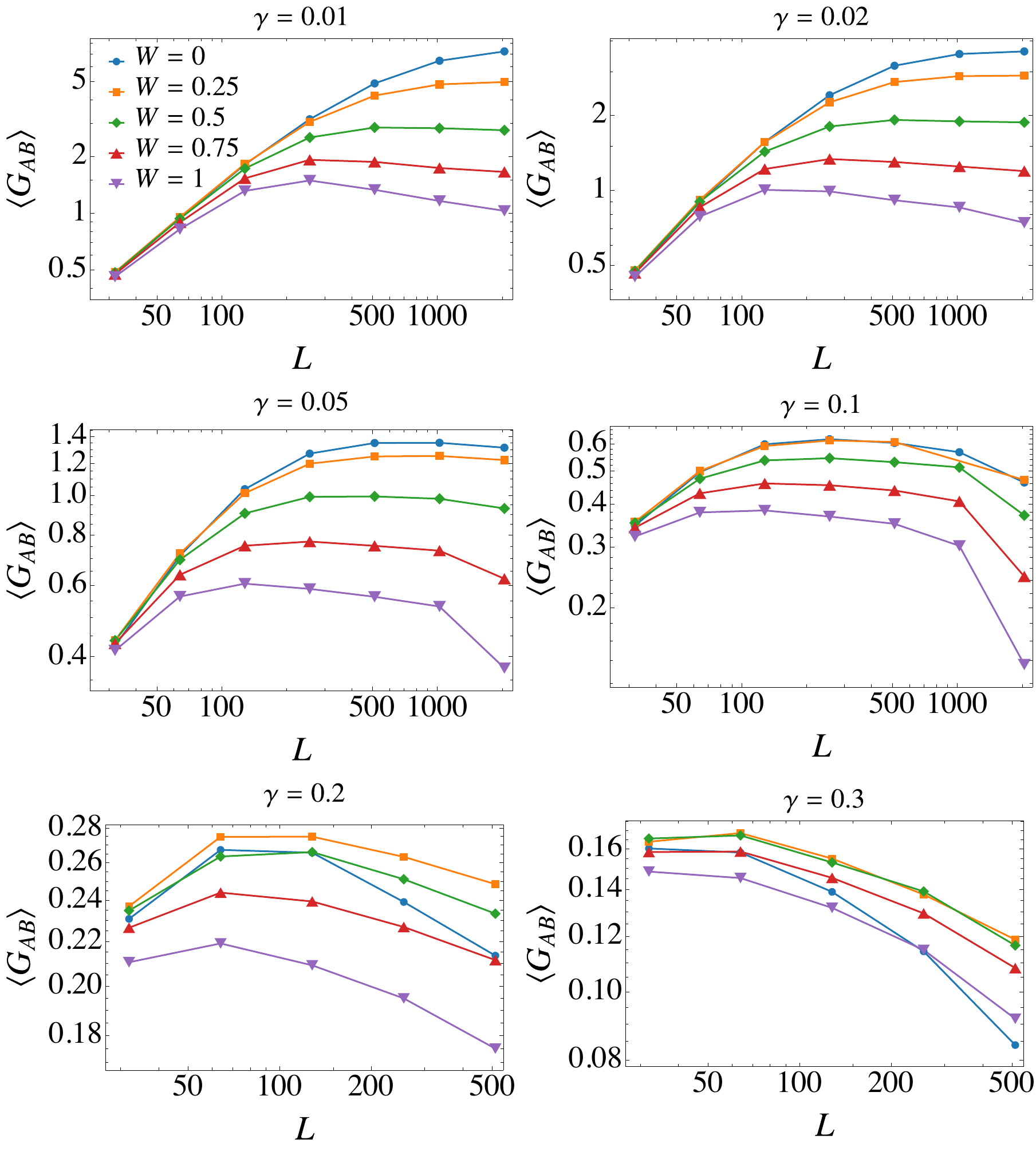}
    \caption{The system size scaling of $G_{AB}$ for 1D disordered monitored fermions for various measurement rates and disorder strengths. The covariance initially increases with system size, but it rapidly reverse the trend to decrease with $L$. }
    \label{fig:GAB-L}
\end{figure}

\begin{figure*}[t]
    \centering
    \includegraphics[width=0.8\linewidth]{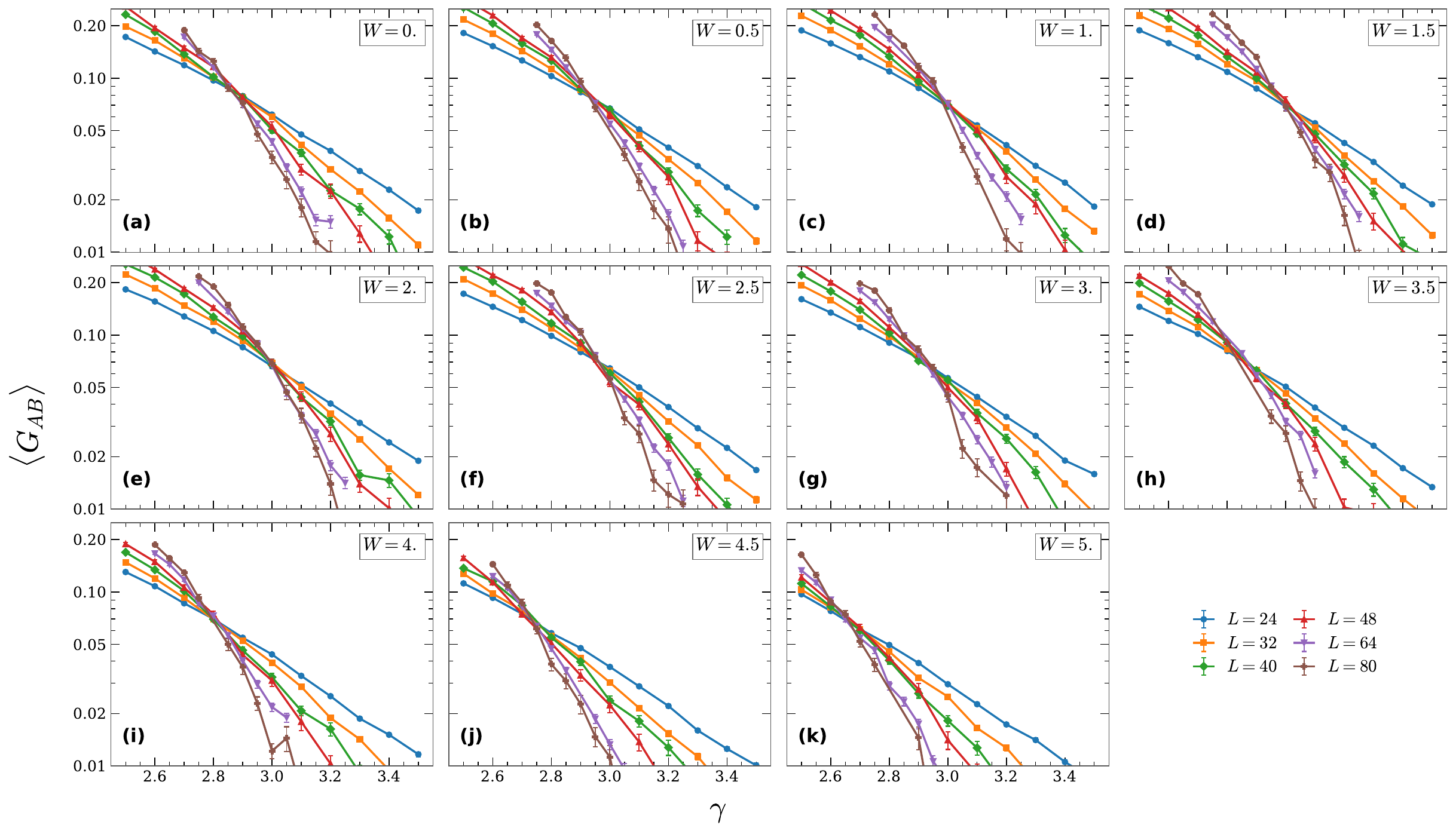}
    \caption{The covariance $G_{AB}$ of 2D monitored fermions at different disorder strengths. The covariance curves for different system sizes cross at approximately the same point even in the presence of on-site disorder potential. This indicates that the MIPT persists in the presence of disorder.}
    \label{fig:GAB-2D}
\end{figure*}
\begin{figure}
    \centering
    \includegraphics[width=0.95\linewidth]
        {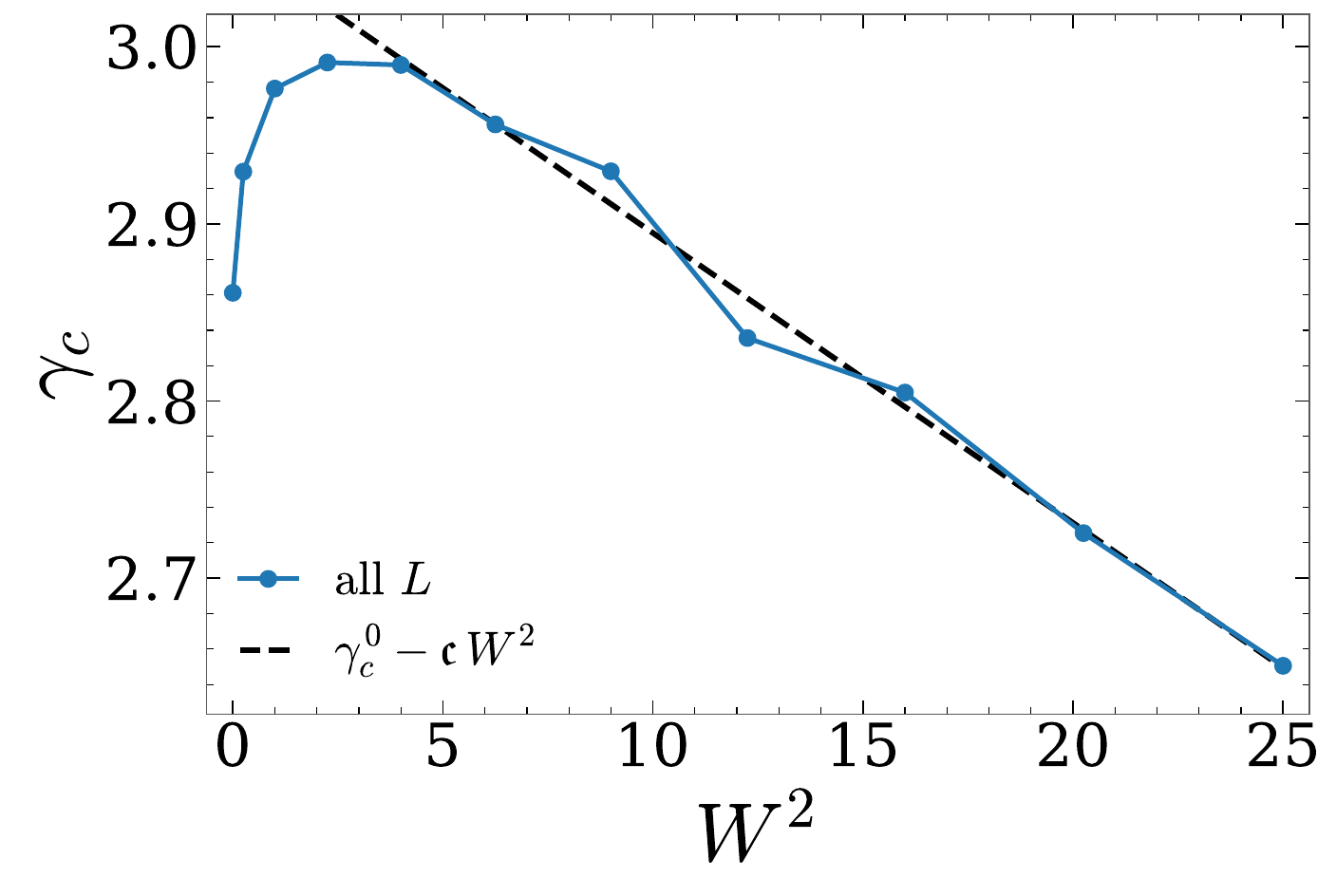}
    \caption{The 2D critical measurement strength $\gamma_c$ as a function of the disorder strength $W$. The value of $\gamma_c$ is obtained by estimating the crossing point of the covariance $G_{AB}$ for different system sizes. The dashed line shows a fit to the data for $W\ge 2$ (beyond the weak disorder symmetry crossover regime) using $\gamma_c = \gamma_c^0 - c W^2$. Here $c$ and $\gamma_c^0$ are fit parameters, and $\gamma_c^0$ can be considered as the clean limit critical measurement strength within the AIII symmetry class, i.e., in the absence of the additional particle-hole symmetry. This result is consistent with the analytical prediction that the critical measurement strength decreases linearly with the disorder-induced decay rate $\gd\propto W^2$.  
    }
    \label{fig:gammac_2d}
\end{figure}

We obtain the numerical results for a 1D fermion chain of length $L$ with periodic boundary conditions. The covariance $G_{AB}$ is evaluated for two equal-size patches,  $A$ and $B$, of length $L_A = L_B = L/4$ with a separation of $d=L/4$ [Fig. \ref{fig:GAB-W}(a)]. In the area-law phase, both the mutual information and covariance $G_{AB}$ decay exponentially with the system size $L$. Conversely, if the system is in the logarithmically entangled critical phase, these quantities saturate to a constant value \cite{Furukawa2009, Calabrese2009}. 

The steady-state covariance $G_{AB}$ is plotted in Figs.~\ref{fig:GAB-W}(b) -- (f) for various disorder strengths. For large measurement rate $\gamma$, $G_{AB}$ decreases with system size, confirming that the system is area-law entangled. For weak measurement rates, $G_{AB}$ initially increases with $L$ for small to moderate system sizes.
A genuine MIPT would manifest as a single, size-independent crossing point in the $G_{AB}$ curves. 
However, Fig.~\ref{fig:GAB-W}(b) reveals that the finite-size crossing between curves of adjacent system sizes is not constant; instead, the critical measurement rate decreases with $L$, approaching zero for the largest system size evaluated.

Interestingly, this drift of the intersection point towards $\gamma=0$ also occurs in disordered fermions. Fig.~\ref{fig:gamma_c_l} tracks the finite-size crossing point, $\gamma_c^L$, as a function of system size. For weak disorder potentials ($W=0.25$ and $0.5$), the behavior closely mirrors that of the clean limit, while the crossing point drifts toward zero more rapidly for stronger disorder ($W=0.75$ and $W=1$).
These results clearly indicate the absence of an MIPT in either clean~\cite{Mirlin_free_1D_2023} or dirty 1D fermions, corroborating our analytical prediction from previous sections. 

Fig. \ref{fig:GAB-L} illustrates the system-size scaling of the covariance $G_{AB}$ for different measurement rate and disorder strength. For small systems, $G_{AB}$ initially increases with $L$; however, it rapidly begins to decay with $L$ once the system is sufficiently large. The overall trend demonstrates 1D free fermions strictly remain in the area-law phase, confirming the absence of a true MIPT in either clean or disordered fermions.

\subsection{Monitored Dirty Fermions in 2D: Disorder Correction to MIPT}

Using the same numerical methods as in the 1D case, we study the quantum projection dynamics of disordered fermions in 2D. Fig.~\ref{fig:GAB-2D} shows the steady-state covariance $G_{AB}$ for 2D monitored fermions,
with different disorder strengths. 
The regions $A$ and $B$ are now strips of size $L\times L/4$ separated by a distance $L/4$.
With this geometry, the covariance $G_{AB}$ decays exponentially with $L$ in the area-law phase, grows linearly with $L$ in the area$\times$log phase, and is independent of $L$ at the transition~\cite{Mirlin_free_2D_2024}.
For 2D clean fermions [Fig.~\ref{fig:GAB-2D}(a), the $G_{AB}$ curves for different system sizes crosses at a common point, reproducing the results of Ref. \cite{Mirlin_free_2D_2024}. Fig.~\ref{fig:GAB-2D}(b)-(k) shows $G_{AB}$ as a function of the measurement rate $\gamma$ for increasing disorder strength. The crossing of the curves for different system sizes persists in the presence of disorder, indicating that the MIPT survives in the presence of disorder. 

Fig.~\ref{fig:gammac_2d} shows the critical measurement rate $\gamma_c$ for different disorder strengths. At weak disorder ($W<2$), $\gamma_c$ increases with the disorder strength $W$, reflecting the change in symmetry induced by the on-site disorder \cite{Mirlin_ineracting_2025}. In the clean limit, the half-filled system on the square lattice possesses an additional effective particle-hole symmetry.  
As a result, the monitored fermions are described by a class BDI NL$\sigma$M when the system is disorder-free. On-site disorder breaks this symmetry, changing the symmetry class of the NL$\sigma$M from BDI to AIII. The increase in $\gamma_c$ at weak disorder $W<2$ reflects this symmetry change.
Our analytical calculation applies to the AIII NL$\sigma$M and therefore does not describe this weak disorder symmetry-crossover regime. When the disorder becomes relatively stronger ($W\ge 2$), $\gamma_c$ decreases with increasing disorder strength, and the change in $\gamma_c$ is proportional to $W^2$. 
For the uniformly distributed disorder used in the numerical simulations, the disorder-induced elastic scattering rate is given by $\gd \propto W^2$. Thus, the numerical results are consistent with the analytical result in Eq.~\eqref{eq:gmc}, which predicts that the critical measurement rate decreases linearly with disorder-induced scattering rate $\gd \propto W^2$.

 \section{Conclusion and Outlook}\label{sec:conclusion}

In this work, we study the interplay between quenched disorder and continuous monitoring of the local particle density in a $d$-dimensional system of noninteracting fermions. We formulate a two-matrix field theory composed of a nonlocal-in-time matrix field 
$\hat{Q}$ that encodes the quenched disorder and a local-in-time matrix field $\hat{q}$ that captures the continuous monitoring. We find that, in the weak-measurement and weak-disorder regime, the long-time, long-distance behaviors of monitored disordered fermions are governed by the Goldstone modes corresponding to simultaneous local-in-time unitary rotations of the two matrix fields, and that the resulting effective theory reduces to the same NL$\sigma$M describing monitored clean fermions, with disorder entering through renormalized model parameters. This finding suggests that disorder does not affect the presence or absence of an MIPT in any dimensions. In particular, in one dimension, our result implies that disorder does not stabilize the critical phase or induce an MIPT, contrary to earlier numerical studies. In $d>1$, our theory predicts that disorder lowers the critical measurement rate (Eq.~\ref{eq:gmc}). Numerical simulations of 1D and 2D lattice models provide supporting evidence for these conclusions: the absence of an MIPT in 1D and the linear decrease of the critical measurement strength with increasing disorder-induced scattering rate. 


Recently, monitoring in quasiperiodic free-fermion system has also been studied \cite{Matsubara2025, Singhaetal2026, matsubara2026}, and an MIPT has also been reported in 1D  \cite{Matsubara2025}. The quasiperiodic potential acts as a kind of structured disorder in modifying the wave function statistics, which does not change the symmetry class relative to the disordered case considered here. Therefore, we expect that it shares similar long-time, long-distance physics with the case of disordered fermions and that the observed MIPT could also be an artifact of finite-size effects.

Another natural next step is to ask how interactions modify the behaviors of disordered monitored fermions considered here. In the clean case~\cite{Foster_interacting_2025,Mirlin_ineracting_2025,Diehl_interacting_fermions_2025,foster2025Maj}, it has been shown that the interactions gap out some of the Goldstone modes present in the noninteracting theory, which are responsible for localization-like corrections that stabilize the area law behavior of monitored fermions in 1D. The interaction-induced mass terms for these modes cut off this ``localization" correction and allow an MIPT to emerge even in 1D~\cite{Foster_interacting_2025,Mirlin_ineracting_2025,tiutiakina2025field}. This effect is analogous to the dephasing of weak localization in conventional disordered systems~\cite{AAK,AAG,Liao_disorder_saddle_point_2017}, where interactions destroy single-particle phase coherence and cut off weak localization corrections, often manifested by the appearance of a mass term in the Cooperon propagator. Given that both clean and disordered noninteracting monitored fermions are described by the same NL$\sigma$M, we expect that interactions would play a similar role in the two cases: they generate a mass term for the soft modes in the current theory and could enable an MIPT in 1D disordered case as well. Detailed investigation of interacting monitored fermions in the presence of disorder is an important direction for future studies.

\textit{Note added.-} Shortly after this work was posted on arXiv, Ref.~\cite{Garcia} appeared, independently studying the one-dimensional case of monitored noninteracting fermions in the presence of disordered and quasiperiodic potentials and reaching the same conclusion that no MIPT occurs in one dimension.

 \section{Acknowledgments}

We thank Jens H. Bardarson for helpful discussions that motivated this work. YL and XZ acknowledge support by the Knut and Alice Wallenberg Foundation Project No.~KAW 2024.0131. Computational resources were provided by the National Academic Infrastructure for Supercomputing in Sweden (NAISS), funded by the Swedish Research Council.

\appendix

\section{Dressed propagators for the Gaussian fluctuations}\label{sec:app-dp}

In this appendix, we present the explicit expressions for the Gaussian action and  the dressed propagators of the Gaussian fluctuations $\delta \hat{q}$ and $\delta \hat{\mathcal{Q}}(0)$ in both the replica-symmetric and replica-asymmetric sectors.

\bigskip

\subsection{Replica-symmetric sector}

The Gaussian action governing the fluctuations in the replica-symmetric sector $\delta \hat{\bar{q}}^{(s)}$ and $\delta \hat{\mathcal{Q}}^{(s)}(0)$ takes the form
\begin{equation}
\begin{aligned}
    &iS^{(2s)}=- R\frac{\pi\nu}{\tilde{\gamma}}\Lambda_{\e}
    \int_{\omega,\mathbf{k}} J^{\msf{q}(s)}_{-\omega,-\mathbf{k}}J^{\msf{q}(s)}_{\omega,\mathbf{k}}
    \\
    &-R\frac{\pi\nu}{\tilde{\gamma}}
    \Lambda_{\e}\!\!\!
    \int_{\omega,\mathbf{k}}\!\!\!
    \begin{bmatrix}
    \delta \bar{q}^{(s)12}_{\omega,\mathbf{k}}
    &
    \delta \mathcal{Q}^{(s)12}_{\ww,\mathbf{k}}(0)
    \end{bmatrix}
    \hat{K}^{(s)}(\kb,\ww)
    \begin{bmatrix}
    \delta \bar{q}^{(s)21}_{-\omega,-\mathbf{k}}
    \\
    \delta \mathcal{Q}^{(s)21}_{-\ww,-\mathbf{k}}(0)
    \end{bmatrix}
    \\
    &+iR\frac{\pi\nu}{\tilde{\gamma}}\Lambda_{\e}
    \int_{\omega,\mathbf{k}}
    J^{\msf{q}(s)}_{-\omega,-\mathbf{k}}
    \left(
    \gamma_{\msf{m}} \delta \bar{q}^{(s)12}_{\omega,\mathbf{k}} 
    +
    \gamma_{\msf{D}} \delta \mathcal{Q}^{(s)12}_{\ww,\mathbf{k}}(0)
    \right.
    \\
    &\qquad \qquad \qquad \left. +
    \gamma_{\msf{m}}\delta  \bar{q}^{(s)21}_{\omega,\mathbf{k}}
    +
    \gamma_{\msf{D}} \delta \mathcal{Q}^{(s)21}_{\omega,\mathbf{k}}(0)
    \right),
\end{aligned}
\label{eqn:S-U1}
\end{equation}
where 
\begin{align}
\begin{aligned}
    &\hat{K}^{(s)}
    =
    \begin{bmatrix}
    K_{mm} & K_{mD}
    \\
    K_{Dm} &K_{DD}
    \end{bmatrix},
    \\
 &K_{mm}(\omega,\mathbf{k})
 =\gm^2
 \left(\frac{\gd}{\gamma_{\msf{m}}}+\theta(\omega,\mathbf{k})-\phi(\omega)\right),
 \\
 &K_{DD}(\omega,\mathbf{k})
 =\gd^2
 \left(\frac{\gm}{\gd}+\theta(\omega,\mathbf{k})-\phi(\omega)\right),
 \\
 &K_{Dm}(\omega,\mathbf{k})=
 K_{mD}(\omega,\mathbf{k})
 =
 \gm\gd
 \left(-1+\theta(\omega,\mathbf{k})-\phi(\omega)\right).
\end{aligned}
\end{align}

The corresponding dressed propagators for the fluctuations $\delta \hat{\bar{q}}^{(s)}$ and $\delta \hat{\mathcal{Q}}^{(s)}(0)$, governed by the action above, are given by
\begin{align}\label{eqn:dres-Prop-1}
\begin{aligned}
    &\braket{
    \delta \bar{q}^{(s)21}_{-\omega,-\mathbf{k}}
    \delta \bar{q}^{(s)12}_{\omega,\mathbf{k}}
    }
    =
    \frac{1}{R\pi \nu \Lambda_{\e}} 
    \frac{1}{\tilde{\gamma}}
    \left(
    \dfrac{1}{\theta(\omega,k)-\phi(\omega)}
    +\frac{\gamma_{\msf{D}}}{\gamma_{\msf{m}}}
    \right),
    \\
    &\braket{
    \delta \mathcal{Q}^{(s)21}_{-\ww,-\mathbf{k}}(0)
    \delta \mathcal{Q}^{(s)12}_{\ww,\mathbf{k}}(0)
    }
        =
    \frac{1}{R\pi \nu \Lambda_{\e}} 
    \frac{1}{\tilde{\gamma}}
    \left(
    \dfrac{1}{\theta(\omega,k)-\phi(\omega)}
    +\frac{\gamma_{\msf{m}}}{\gamma_{\msf{D}}}
    \right),
    \\
    &\braket{
    \delta q^{(s)21}_{-\ww,-\mathbf{k}}
    \delta \mathcal{Q}^{(s)12}_{\ww,\mathbf{k}}(0)
    }
    =
    \braket{
    \delta \mathcal{Q}^{(s)21}_{\ww,\mathbf{k}}(0)
    \delta q^{(s)12}_{\ww,\mathbf{k}}
    }
        \\
    &=
    \frac{1}{R\pi \nu \Lambda_{\e} } 
    \frac{1}{\tilde{\gamma}}
    \left(
    \dfrac{1}{\theta(\omega,k)-\phi(\omega)}
    -1
    \right).
\end{aligned}
\end{align}

\subsection{Replica-asymmetric sector}

In the replica-asymmetric sector, the Gaussian fluctuations are further decomposed into $Y^{(\msf{m})}$ and $\mathcal{Y}^{(\msf{D})}$, corresponding to the $\tau^2$ components of the fluctuations $\delta \hat{\bar{q}}^{(a)}$ and $\delta \hat{\mathcal{Q}}^{(a)}(0)$, respectively, as well as 
$X^{(\msf{m})}$ and $\mathcal{X}^{(\msf{D})}$, which are the corresponding $\tau^1$ components.
They are governed by the action
\begin{widetext}
\begin{equation}
\begin{aligned}
    S^{(2a)}=&S_X+S_Y+S_{XY}+S_{XYJ}+S_{J0},
    \\
    iS_{X}
    =&
    -\frac{\pi\nu}{\tilde{\gamma}}
    \Lambda_{\e}
    \int_{\omega,\mathbf{k}}
    \tr_R
    \left\lbrace
    \begin{bmatrix}
    \hat{X}^{(\msf{m})}_{\omega,\mathbf{k}}
    &
    \hat{\mathcal{X}}^{(\msf{D})}_{\ww,\mathbf{k}}
    \end{bmatrix}
    \begin{bmatrix}
    K_{mm}^{(X)}(\omega,\mathbf{k}) & K_{mD}'(\omega,\mathbf{k})
    \\
    K_{Dm}'(\omega,\mathbf{k}) &K_{DD}'(\omega,\mathbf{k})
    \end{bmatrix}
    \begin{bmatrix}
    \hat{X}^{(\msf{m})}_{-\omega,-\mathbf{k}}
    &
    \hat{\mathcal{X}}^{(\msf{D})}_{-\ww,-\mathbf{k}}
    \end{bmatrix}
    \right\rbrace,
    \\
    iS_{XY}  =&+2i\frac{\pi\nu}{\tilde{\gamma}}\Lambda_{\e}
    \int_{\omega,\mathbf{k}}
    \phi(\ww)
    \tr_R\left[
    \left(
    \gm \hat{X}^{(\msf{m})}_{\omega,\mathbf{k}}
    +
    \gd
    \hat{\mathcal{X}}^{(\msf{D})}_{\ww,\mathbf{k}}
    \right)
    \left(
    \gm \hat{Y}^{(\msf{m})}_{-\omega,-\mathbf{k}}
    +
    \gd
    \hat{\mathcal{Y}}^{(\msf{D})}_{-\ww,-\mathbf{k}}
    \right)
    \right],
    \\
    iS_{XYJ}  
    =&
    -2i\frac{\pi\nu}{\tilde{\gamma}}
    \Lambda_{\e}
    \int_{\omega,\mathbf{k}}
    \tr_R\left\lbrace
    \hat{\mathcal{J}}^{\msf{q}(a)}_{-\omega,-\mathbf{k}}
    \left[
    i\phi(\ww)
    \left(
    \gm \hat{Y}^{(\msf{m})}_{\omega,\mathbf{k}}
    +
    \gd
    \hat{\mathcal{Y}}^{(\msf{D})}_{\ww,\mathbf{k}}
    \right)
    -
    \left[1-\theta(\omega,k)\right]
    \left(
    \gm \hat{X}^{(\msf{m})}_{\omega,\mathbf{k}}
    +
    \gd
    \hat{\mathcal{X}}^{(\msf{D})}_{\ww,\mathbf{k}}
    \right)
    \right]
    \right\rbrace,
    \\
    iS_{J0}=  &-
    \frac{\pi\nu}{\tilde{\gamma}}
    \Lambda_{\e}
    \int_{\omega,\mathbf{k}}
    \left[1-\theta(\omega,k)\right]
    \tr_R\left(
    \hat{\mathcal{J}}^{\msf{q}(a)}_{-\omega,-\mathbf{k}}
    \hat{\mathcal{J}}^{\msf{q}(a)}_{\omega,\mathbf{k}}
    \right),
\end{aligned}
\label{eqn:S-SUR}
\end{equation}
\end{widetext}
where
\begin{align}
\begin{aligned}
 &K_{mm}^{(X/Y)}(\omega,\mathbf{k})
 =\gm^2
 \left(
 -1 \mp \frac{\tilde{\gamma}}{\gamma_{\msf{m}}}
 +\theta(\omega,\mathbf{k})\right),
 \\
 &K_{DD}'(\omega,\mathbf{k})
 =\gd^2
 \left(\frac{\gm}{\gd}+\theta(\omega,\mathbf{k})\right),
 \\
 &K_{Dm}'(\omega,\mathbf{k})=
 K_{mD}'(\omega,\mathbf{k})
 =
 \gm\gd
 \left(-1+\theta(\omega,\mathbf{k})\right).
\end{aligned}
\end{align}
The action $S_Y$ has the same quadratic form as $S_X$,
 upon replacing $\hat{X}^{(\msf{m})} \to \hat{Y}^{(\msf{m})}$,  $\hat{\mathcal{X}}^{(\msf{D})}\to\hat{\mathcal{Y}}^{(\msf{D})}$, and $K_{mm}^{(X)} \to K_{mm}^{(Y)}$.

The propagators for $Y^{(\msf{m})}$ and $\mathcal{Y}^{(\msf{D})}$ arising from $S_Y$ in Eq.~\ref{eqn:S-SUR} are given by
\begin{align}\label{eqn:Y-Prop}
\begin{aligned}
    &\braket{
    Y_{-\omega,-\mathbf{k}}^{(\msf{m})ij}
    Y_{\omega,\mathbf{k}}^{(\msf{m})uv}
    }
    =
    \frac{P_{ij,uv}}{2 \pi \nu \Lambda_{\e}} 
    \frac{1}{\tilde{\gamma}}
    \left(
    \dfrac{1}{\theta(\omega,k)}
    +\frac{\gamma_{\msf{D}}}{\gamma_{\msf{m}}}
    \right),
    \\
    &\braket{
    \mathcal{Y}^{(\msf{D})ij}_{-\ww,-\mathbf{k}}
    \mathcal{Y}^{(\msf{D})uv}_{\ww,\mathbf{k}}
    }
    =
    \frac{P_{ij,uv}}{2 \pi \nu \Lambda_{\e}} 
    \frac{1}{\tilde{\gamma}}
    \left(
    \dfrac{1}{\theta(\omega,k)}
    +\frac{\gamma_{\msf{m}}}{\gamma_{\msf{D}}}
    \right),
    \\
    &\braket{
   Y_{-\omega,-\mathbf{k}}^{(\msf{m})ij}
    \mathcal{Y}^{(\msf{D})uv}_{\ww,\mathbf{k}}
    }
    =
    \frac{P_{ij,uv}}{2 \pi \nu \Lambda_{\e}} 
    \frac{1}{\tilde{\gamma}}
    \left(
    \dfrac{1}{\theta(\omega,k)}
    -1
    \right).
\end{aligned}
\end{align}
Similarly, the propagators for $X^{(\msf{m})}$ and $\mathcal{X}^{(\msf{D})}$ arising from $S_X$ are
\begin{align}\label{eqn:X-Prop}
\begin{aligned}
    &\braket{
    X_{-\omega,-\mathbf{k}}^{(\msf{m})ij}
    X_{\omega,\mathbf{k}}^{(\msf{m})uv}
    }
    =
    -
    \frac{P_{ij,uv}}{2 \pi \nu \Lambda_{\e}} 
    \frac{1}{2\gamma_{\msf{m}}} 
    \dfrac{1+\frac{\gd}{\gm}\theta(\omega,k)}{1-\frac{\gm-\gd}{2\gm}\theta(\omega,k)},
    \\
    &\braket{
    \mathcal{X}^{(\msf{D})ij}_{-\omega,-\mathbf{k}}
    \mathcal{X}^{(\msf{D})uv}_{\ww,\mathbf{k}}
    }
    =
    \frac{P_{ij,uv}}{2 \pi \nu \Lambda_{\e}}
    \frac{(\gd+2\gm)}{2\gd\gm}
    \dfrac{1-\frac{\gm}{\gd+2\gm}\theta(\omega,k)}{1-\frac{\gm-\gd}{2\gm}\theta(\omega,k)},
    \\
    &\braket{
   X_{-\omega,-\mathbf{k}}^{(\msf{m})ij}
    \mathcal{X}^{(\msf{D})uv}_{\ww,\mathbf{k}}
    }
    =
    -\frac{P_{ij,uv}}{2 \pi \nu \Lambda_{\e}}
    \frac{1}{2\gamma_{\msf{m}}}
    \dfrac{1-\theta(\omega,k)}{1-\frac{\gm-\gd}{2\gm}\theta(\omega,k)}.
\end{aligned}
\end{align}

\section{Alternative treatment of the Gaussian fluctuations}\label{sec:app-dp-2}

In this appendix, we provide an alternative Gaussian approximation analysis, retaining all fluctuation modes that are massless in the measurement-only or disorder-only theory. In particular, we include all components of $\delta \hat{Q}_{t_1,t_2}$, including the off-diagonal in time modes with $t_1\neq t_2$, which are massive in the presence of measurement and decouple from all massless modes and source fields at quadratic order. This analysis clarifies the full structure of Gaussian fluctuations, especially the role of the massive and nonlocal-in-time components of the $\hat{Q}$ matrix field.
For completeness, we list below all propagators in both sectors. The $\delta \hat{
\bar{q}}$ propagators are identical to those in Appendix~\ref{sec:app-dp}, since $\delta \hat{
\bar{q}}$ is local in time. By contrast,  $\delta \hat{
Q}$ fluctuations carry additional nonlocal-in-time structure, and their propagators reproduce those of the time-diagonal component $\delta \hat{\mathcal{Q}}$ given in Appendix~\ref{sec:app-dp}.

\subsection{Replica-symmetric sector}

The action for Gaussian fluctuations in the replica-symmetric sector $\delta \hat{\bar{q}}^{(s)}$ and $\delta \hat{Q}^{(s)}$  takes the form
\begin{equation}
\begin{aligned}
    iS^{(s)}
    =
    &-R\pi\nu\gamma_{\msf{m}}
    \int_{\e} 
    \int_{\omega,\mathbf{k}}
    \delta \bar{q}^{(s)12}_{\omega,\mathbf{k}}
    \delta \bar{q}^{(s)21}_{-\omega,-\mathbf{k}}
    \\
    &-R\pi\nu\gamma_{\msf{D}}
    \int_{\varepsilon,\omega,\mathbf{k}}
    \delta Q^{(s)12}_{\e,\e-\ww,\mathbf{k}}
    \delta Q^{(s)21}_{\e-\ww,\e,-\mathbf{k}}
    \\
    &-R\frac{\pi\nu}{\tilde{\gamma}}
    \int_{\varepsilon,\omega,\mathbf{k}}
    \left[-1+\theta(\omega,k)-\phi(\omega)\right]
    \\
    &\qquad \times
    \left[
    \gamma_{\msf{m}} \delta \bar{q}^{(s)12}_{\omega,\mathbf{k}} 
    +
    \gamma_{\msf{D}} \delta Q^{(s)12}_{\e,\e-\ww,\mathbf{k}}
    +
    i{J}^{\msf{q}(s)}_{\omega,\mathbf{k}}
    \right]
    \\
    &\qquad \times
    \left[
    \gamma_{\msf{m}}\delta  \bar{q}^{(s)21}_{-\omega,-\mathbf{k}}
    +
    \gamma_{\msf{D}} \delta Q^{(s)21}_{\varepsilon-\omega,\varepsilon,-\mathbf{k}}
    +
    iJ^{\msf{q}(s)}_{-\omega,-\mathbf{k}}
    \right],
\end{aligned}
\label{eqS:S-U1}
\end{equation}
 
\begin{widetext}
As in the main text, we keep only those fluctuations that are massless in the measurement-only or disorder-only limits. In the presence of both disorder and measurement, these fluctuations appear massive in their respective bare actions:
\begin{align}
\begin{aligned}\label{eqn:bare-Prop}
    &\braket{
    \delta \bar{q}^{(s)21}_{-\omega,-\mathbf{k}}
    \delta \bar{q}^{(s)12}_{\omega',\mathbf{k}'}
    }_0
    =
    \frac{1}{R\pi \nu \Lambda_{\e}} 
    \frac{\tilde{\gamma}}{\gamma_{\msf{m}}^2} 
    \dfrac{1}{\frac{{\gamma}_D}{\gamma_{\msf{m}}}+\theta(\omega,k)-\phi(\omega)},
    \\
    &\braket{
    \delta Q^{(s)21}_{\e-\ww,\e,-\mathbf{k}}
    \delta Q^{(s)12}_{\e',\e'-\ww,\mathbf{k}}
    }_0
    =
    \frac{1}{R\pi \nu} \frac{\tilde{\gamma}}{\gamma_{\msf{D}}^2}  
    \dfrac{1}{\frac{{\gamma}_m}{\gamma_{\msf{D}}}+\theta(\omega,k)-\phi(\omega)}
    \delta_{\e,\e'}.
\end{aligned}
\end{align}
The interaction between matrix fields $\delta \hat{Q}^{(s)}$ and $\delta \hat{\bar{q}}^{(s)}$ leads to the following dressed propagators
\begin{align}\label{eqn:dres-Prop-2}
\begin{aligned}
    &\braket{
    \delta \bar{q}^{(s)21}_{-\omega,-\mathbf{k}}
    \delta \bar{q}^{(s)12}_{\omega,\mathbf{k}}
    }
    =
    \frac{1}{R\pi \nu \Lambda_{\e}} 
    \frac{1}{\tilde{\gamma}}
    \left(
    \dfrac{1}{\theta(\omega,k)-\phi(\omega)}
    +\frac{\gamma_{\msf{D}}}{\gamma_{\msf{m}}}
    \right),
    \\
    &\braket{
    \delta Q^{(s)21}_{\e-\ww,\e,-\mathbf{k}}
    \delta Q^{(s)12}_{\e',\e'-\ww,\mathbf{k}}
    }
    =
    \frac{1}{R \pi \nu} \frac{\tilde{\gamma}}{\gamma_{\msf{D}}^2}
    \dfrac{1}{\frac{\gamma_{\msf{m}}}{\gamma_{\msf{D}}}+\theta(\omega,k)-\phi(\omega)}
    \left(
    \delta_{\e,\e'}
    +
    \frac{1}{\Lambda_{\e}}
    \frac{\gamma_{\msf{m}}\gamma_{\msf{D}}}{\tilde{\gamma}^2}
    \dfrac{\left(-1+\theta(\omega,k)-\phi(\omega)\right)^2}{\theta(\omega,k)-\phi(\omega)}
    \right),
    \\
    &\braket{
    \delta q^{(s)21}_{-\ww,-\mathbf{k}}
    \delta Q^{(s)12}_{\e,\e-\ww,\mathbf{k}}
    }
    =
    \braket{
    \delta Q^{(s)21}_{\e-\ww,\e,\mathbf{k}}
    \delta q^{(s)12}_{\ww,\mathbf{k}}
    }
    =
    \frac{1}{R\pi \nu \Lambda_{\e}}
    \frac{1}{\tilde{\gamma}} 
    \left(
    \dfrac{1}{\theta(\omega,k)-\phi(\omega)}
    -1
    \right).
\end{aligned}
\end{align}
From these results, we can reproduce the propagator for the mixed modes $\delta \hat{\mathcal{Q}}^{(+)}$ in Eq.~\ref{eqn:dress-Prop-sym}, which in turn determines the average-density response function $\Pi^{(s)}(\kb,\ww)$.

 \subsection{Replica-asymmetric sector}

The action governing the Gaussian fluctuations in the replica-asymmetric sector assumes the form
\begin{equation}
\begin{aligned}
    S^{(a)}=&S_X+S_Y+S_{XY}+S_{XYJ}+S_{J},
    \\
    iS_X
    =&-\frac{\pi\nu}{\tilde{\gamma}}
    \int_{\e} 
    \int_{\omega,\mathbf{k}}
    \tr_R\left\lbrace
    \gm^2
    \hat{X}^{(\msf{m})}_{\omega,\mathbf{k}}
     \left[-1-\frac{\tilde{\gamma}}{\gm}
     +\theta(\omega,k)\right]
    \hat{X}^{(\msf{m})}_{-\omega,-\mathbf{k}}
    +
    \gd^2
    \hat{X}^{(\msf{D})}_{\e,\e-\ww,\mathbf{k}}
     \left[-1+\frac{\tilde{\gamma}}{\gd}+\theta(\omega,k)\right]
    \hat{X}^{(\msf{D})}_{\e-\ww,\e,-\mathbf{k}}
    \right.
    \\
    &\left.+
    2\gm\gd
    \hat{X}^{(\msf{m})}_{\omega,\mathbf{k}}
     \left[-1+\theta(\omega,k)\right]
    \hat{X}^{(\msf{D})}_{\e-\ww,\e,-\mathbf{k}}
    \right\rbrace,
    \\
    iS_Y
    =&-\frac{\pi\nu}{\tilde{\gamma}}
    \int_{\e} 
    \int_{\omega,\mathbf{k}}
    \tr_R\left\lbrace
    \gm^2
    \hat{Y}^{(\msf{m})}_{\omega,\mathbf{k}}
     \left[-1+\frac{\tilde{\gamma}}{\gm}
     +\theta(\omega,k)\right]
    \hat{Y}^{(\msf{m})}_{-\omega,-\mathbf{k}}
    +
    \gd^2
    \hat{Y}^{(\msf{D})}_{\e,\e-\ww,\mathbf{k}}
     \left[-1+\frac{\tilde{\gamma}}{\gd}+\theta(\omega,k)\right]
    \hat{Y}^{(\msf{D})}_{\e-\ww,\e,-\mathbf{k}}
    \right.
    \\
    &\left.+
    2\gm\gd
    \hat{Y}^{(\msf{m})}_{\omega,\mathbf{k}}
     \left[-1+\theta(\omega,k)\right]
    \hat{Y}^{(\msf{D})}_{\e-\ww,\e,-\mathbf{k}}
    \right\rbrace,
    \\
    iS_{XY}  =&+2i\frac{\pi\nu}{\tilde{\gamma}}
    \int_{\e} 
    \int_{\omega,\mathbf{k}}
    \phi(\ww)
    \tr_R\left\lbrace
    \left(
    \gm \hat{X}^{(\msf{m})}_{\omega,\mathbf{k}}
    +
    \gd
    \hat{X}^{(\msf{D})}_{\e,\e-\ww,\mathbf{k}}
    \right)
    \left(
    \gm \hat{Y}^{(\msf{m})}_{-\omega,-\mathbf{k}}
    +
    \gd
    \hat{Y}^{(\msf{D})}_{\e-\ww,\e,-\mathbf{k}}
    \right)
    \right\rbrace,
    \\
    iS_{XYJ}  
    =&
    -i2\frac{\pi\nu}{\tilde{\gamma}}
    \int_{\e} 
    \int_{\omega,\mathbf{k}}
    \tr_R\left\lbrace
    \hat{\mathcal{J}}^{\msf{q}(a)}_{-\omega,-\mathbf{k}}
    \left\lbrace
    i\phi(\ww)
    \left(
    \gm \hat{Y}^{(\msf{m})}_{\omega,\mathbf{k}}
    +
    \gd
    \hat{Y}^{(\msf{D})}_{\e,\e-\ww,\mathbf{k}}
    \right)
    +
    \left[-1+\theta(\omega,k)\right]
    \left(
    \gm \hat{X}^{(\msf{m})}_{\omega,\mathbf{k}}
    +
    \gd
    \hat{X}^{(\msf{D})}_{\e,\e-\ww,\mathbf{k}}
    \right)
    \right\rbrace
    \right\rbrace
    \\
    iS_{J}=  &-
    \frac{\pi\nu}{\tilde{\gamma}}
    \int_{\e} 
    \int_{\omega,\mathbf{k}}
    \left[1-\theta(\omega,k)\right]
    \tr_R\left(
    \hat{\mathcal{J}}^{\msf{q}(a)}_{-\omega,-\mathbf{k}}
    \hat{\mathcal{J}}^{\msf{q}(a)}_{\omega,\mathbf{k}}
    \right).
\end{aligned}
\label{eqn:S-SUR-2}
\end{equation}
Here we have defined, as before, $X^{(\msf{D})}$ and $Y^{(\msf{D})}$ as the $\hat{\tau}^1$ and $\tau^2$ components of $\delta \hat{Q}^{(a)}$:
\begin{align}
\begin{aligned}
    &\hat{X}^{(\msf{D})}=(\delta \hat{Q}^{(a)12}+\delta \hat{Q}^{(a)21})/2,
    \qquad
    \hat{Y}^{(\msf{D})}=i(\delta \hat{Q}^{(a)12}-\delta \hat{Q}^{(a)21})/2.
\end{aligned}
\end{align}

The $Y$ propagators arising solely from the action $S_Y$ are given by
\begin{align}\label{Aeqn:Y-Prop}
\begin{aligned}
    &\braket{
    Y_{-\omega,-\mathbf{k}}^{(\msf{m})ij}
    Y_{\omega,\mathbf{k}}^{(\msf{m})uv}
    }
    =
    \frac{P_{ij,uv}}{2 \pi \nu \Lambda_{\e}}
    \frac{1}{\tilde{\gamma}} 
    \left(
    \dfrac{1}{\theta(\omega,k)}
    +\frac{\gamma_{\msf{D}}}{\gamma_{\msf{m}}}
    \right),
    \\
    &\braket{
    Y^{(\msf{D})ij}_{\e-\ww,\e,-\mathbf{k}}
    Y^{(\msf{D})uv}_{\e',\e'-\ww,\mathbf{k}}
    }
    =
     \frac{P_{ij,uv}}{2 \pi \nu}
    \dfrac{\tilde{\gamma}/\gamma_{\msf{D}}^2}{\frac{\gamma_{\msf{m}}}{\gamma_{\msf{D}}}+\theta(\omega,k)}
    \left(
    \delta_{\e,\e'}
    +\frac{1}{\Lambda_\e}
    \frac{\gamma_{\msf{m}}\gamma_{\msf{D}}}{\tilde{\gamma}^2}
    \dfrac{\left(\theta(\omega,k)-1\right)^2}{\theta(\omega,k)}
    \right),
    \\
    &\braket{
   Y_{-\omega,-\mathbf{k}}^{(\msf{m})ij}
    Y^{(\msf{D})uv}_{\e,\e-\ww,\mathbf{k}}
    }
    =
    \frac{P_{ij,uv}}{2 \pi \nu \Lambda_{\e}}
    \frac{1}{\tilde{\gamma}}
    \left(
    \dfrac{1}{\theta(\omega,k)}
    -1
    \right).
\end{aligned}
\end{align}
Similarly, the $X$ propagators arising from the action $S_X$ take  the form
\begin{align}\label{Aeqn:X-Prop}
\begin{aligned}
    &\braket{
    X_{-\omega,-\mathbf{k}}^{(\msf{m})ij}
    X_{\omega,\mathbf{k}}^{(\msf{m})uv}
    }
    =
    -
    \frac{P_{ij,uv}}{2 \pi \nu \Lambda_{\e}}
    \frac{1}{2\gamma_{\msf{m}}} 
    \dfrac{1+\frac{\gd}{\gm}\theta(\omega,k)}{1-\frac{\gm-\gd}{2\gm}\theta(\omega,k)},
    \\
    &\braket{
    X^{(\msf{D})ij}_{\e-\ww,\e,-\mathbf{k}}
    X^{(\msf{D})uv}_{\e',\e'-\ww,\mathbf{k}}
    }
    =
    \frac{P_{ij,uv}}{2 \pi \nu}
    \dfrac{\tilde{\gamma}/\gamma_{\msf{D}}^2}{\frac{\gamma_{\msf{m}}}{\gamma_{\msf{D}}}+\theta(\omega,k)}
    \left(
    \delta_{\e,\e'}
    -
    \frac{1}{\Lambda_\e}
    \frac{\gamma_{\msf{D}}}{2\tilde{\gamma}}
    \dfrac{\left(1-\theta(\omega,k)\right)^2}{1-\frac{\gm-\gd}{2\gm}\theta(\omega,k)}
    \right),
    \\
    &\braket{
   X_{-\omega,-\mathbf{k}}^{(\msf{m})ij}
    X^{(\msf{D})uv}_{\e,\e-\ww,\mathbf{k}}
    }
    =
    -
    \frac{P_{ij,uv}}{2 \pi \nu \Lambda_{\e}}
    \frac{1}{2\gamma_{\msf{m}}} 
    \dfrac{\left(1-\theta(\omega,k)\right)}{1-\frac{\gm-\gd}{2\gm}\theta(\omega,k)}.
\end{aligned}
\end{align}
Using these expressions for the dressed propagators, we can rederive the propagators for $\hat{Y}^{(+)}$ and $\hat{X}^{(+)}$ in Eq.~\ref{eqn:dres-Prop}. The dressed propagators can then be used to obtain the connected particle density correlation function $C(\rb,\rb';t,t')$.

\end{widetext}

\section{Derivation of the NL$\sigma$M}\label{sec:app-nlsm}

In this appendix, we outline the detailed derivation of the NL$\sigma$M in Eq.~\ref{eq:SGM-F2} describing the Goldstone modes in the current theory of monitored disordered fermions.

\subsection{Effective action for the Goldstone modes}\label{sec:app-nlsm-gm}

We first derive the action for the Goldstone modes only, and address the renormalization effect of the massive modes and the coupling to the source field in later sections. 
We start from Eq.~\ref{eq:SGM-1}, which can be rewritten as
\begin{equation}\label{eq:SGM-2}
\begin{aligned}
    &iS_{gm}
    =
    \Tr\ln\left[ 1+i \hat{G}_{\msf{sp}} \hat{\mathcal{R}}^{-1} \partial_t \hat{\mathcal{R}}+ i \hat{G}_{\msf{sp}} \hat{\mathcal{R}}^{-1} \vex{v}\cdot \nabla \hat{\mathcal{R}}\right],
\end{aligned}
\end{equation}
where $\hat{G}_{\msf{sp}}^{-1}=(\hat{G}_0^{-1}+i\tilde{\gamma} \hat{\tau}^3)$.
 Performing the expansion of the action above, the leading order term is $i\Tr\left[\hat{G}_{\msf{sp}} \hat{\mathcal{R}}^{-1} \partial_t \hat{\mathcal{R}}\right]$, which after applying the saddle point equation $\sum_{\kb,\e} \hat{G}_{\msf{sp}}(\kb,\e)=-\frac{i}{2}\left(\frac{\Lambda_k}{2\pi}\right)^d\hat{\tau}^3$ reduces to
\begin{equation}\label{eq:SGM-3}
\begin{aligned}
    &
    i\Tr\left[ \hat{G}_{\msf{sp}} \hat{\mathcal{R}}^{-1} \partial_t \hat{\mathcal{R}}\right]
     =
     \frac{1}{2}\left(\frac{\Lambda_k}{2\pi}\right)^d
     \Tr\left[\hat{\tau}^3 \hat{\mathcal{R}}^{-1} \partial_t \hat{\mathcal{R}}\right]
     \\
     &=
     \frac{1}{2}\left(\frac{\Lambda_k}{2\pi}\right)^d
     \left[R\tr_K\left(\hat{\tau}^3 \hat{\mathcal{R}}^{-1}_0 \partial_t \hat{\mathcal{R}}_0\right)
     +
     \Tr \left(\hat{\mathcal{Q}}_0 \hat{\mathcal{R}}^{-1}_\Theta \partial_t \hat{\mathcal{R}}_\Theta\right)     \right].
\end{aligned}
\end{equation}
Here $\hat{\mathcal{Q}}_0=\hat{\mathcal{R}}_0\hat{\tau}^3 \hat{\mathcal{R}}_0^{-1} \in \mathrm{U}(2)/\mathrm{U}(1)\times \mathrm{U}(1)$. 
The second term in the last equality vanishes because $\tr_R(\hat{\Theta})=0$. We are therefore left with the first term, which is associated with the soft modes in the replica-symmetric sector.

For the term quadratic in $\vex{v}\cdot\nabla$ 
in the $\Tr\ln$ expansion, one can use $\sum_{\kb,\e} G_{\msf{sp}}^R(\kb,\e)G_{\msf{sp}}^A(\kb,\e)=\frac{1}{2\tilde{\gamma}}\left(\frac{\Lambda_k}{2\pi}\right)^d$, which leads to
\begin{equation}\label{eq:SGM-4}
\begin{aligned}
    &
    \frac{1}{2}\Tr\left[\hat{G}_{\msf{sp}} \left(\hat{\mathcal{R}}^{-1} \vex{v}\cdot \nabla \hat{\mathcal{R}} \right)\hat{G}_{\msf{sp}} \left(\hat{\mathcal{R}}^{-1} \vex{v}\cdot \nabla \hat{\mathcal{R}} \right)\right]
     \\
     &=\!\!
     \left(\frac{\Lambda_k}{2\pi}\right)^d\!\! \frac{D}{4}\Tr\left[(1+\hat{\tau}^3) \left(\hat{\mathcal{R}}^{-1}  \nabla \hat{\mathcal{R}} \right)(1-\hat{\tau}^3) \left(\hat{\mathcal{R}}^{-1}  \nabla \hat{\mathcal{R}} \right)\right]
     \\
     &=\!\!-\left(\frac{\Lambda_k}{2\pi}\right)^d\!\!\frac{ D}{8}
     \left[ R
     \tr_K\left[(\nabla \hat{\mathcal{Q}}_0)^2\right]+2\bar{g}[\hat{\mathcal{\mathcal{Q}}}_0]\tr_{R}\left(\nabla \hat{U}^{-1} \nabla \hat{U} \right)
     \right].
\end{aligned}
\end{equation}
Here $\hat{U}=\exp(i\hat{\Theta})\in \rm{SU}(R)$ and $\bar{g}[\hat{\mathcal{Q}}_0]$ is defined as
 \begin{align}
    \bar{g}[\hat{\mathcal{Q}}_0]\equiv \frac{1}{4}\tr_K\left[(1+\hat{\mathcal{Q}}_0)\hat{\tau}^1(1-\hat{\mathcal{Q}}_0)\hat{\tau}^1 \right].
 \end{align}
 We note that the cross term involving both $\nabla \hat{\mathcal{R}}_\Theta$ and $\nabla \hat{\mathcal{R}}_0$ vanishes because $\tr_R(\hat{\Theta})=0$. 
 
Similarly, the term quadratic in $\partial_t$ in the $\Tr\ln$ expansion gives 
 \begin{equation}\label{eq:SGM-5}
\begin{aligned}
    &
    \frac{1}{2}\Tr\left[\hat{G}_{\msf{sp}} \left(\hat{\mathcal{R}}^{-1} \partial_t \hat{\mathcal{R}} \right)\hat{G}_{\msf{sp}} \left(\hat{\mathcal{R}}^{-1} \partial_t \hat{\mathcal{R}} \right)\right]
     \\
     &=-\left(\frac{\Lambda_k}{2\pi}\right)^d\frac{\bar{g}[\hat{\mathcal{Q}}_0]}{8\tilde{\gamma}}
     \tr_{R}\left(\partial_t \hat{U}^{-1} \partial_t \hat{U}\right).
\end{aligned}
\end{equation}
Here we have neglected a term  quadratic in $\partial_t \hat{\mathcal{R}}_0$, as it is of higher order in the gradient expansion than the term in Eq.~\ref{eq:SGM-3}.

Combining all these terms, we arrive at the effective action for the Goldstone modes $\hat{U}\in \mathrm{SU}(R)$ in the replica-asymmetric  sector and $\hat{\mathcal{Q}}_0 \in \mathrm{U}(2)/\mathrm{U}(1)\times \mathrm{U}(1)$ in the replica-symmetric  sector:
 \begin{equation}\label{eq:SGM-F1}
\begin{aligned}
    &iS_{gm}=\left(\frac{\Lambda_k}{2\pi}\right)^d
     \frac{R}{2}\left[
     \tr_K\left(\hat{\tau}^3 \hat{\mathcal{R}}^{-1}_0 \partial_t \hat{\mathcal{R}}_0\right)
    -\frac{ D}{4}
     \tr_K\left[(\nabla \hat{\mathcal{Q}}_0)^2\right]
     \right]
    \\
    &
    -\left(\frac{\Lambda_k}{2\pi}\right)^d\!\!\frac{\bar{g}[\hat{\mathcal{Q}}_0]}{8\tilde{\gamma}}
    \left[
    \tr_{R}\left(\partial_t \hat{U}^{-1} \partial_t \hat{U}\right)
    +
    \frac{v^2}{d}\tr_{R}\left(\nabla \hat{U}^{-1} \nabla \hat{U}\right)
    \right].
\end{aligned}
\end{equation}
Here the overall factor $\bar{g}[\hat{\mathcal{Q}}_0]$ in the second line leads to the coupling between the soft modes in the replica-asymmetric and replica-symmetric sectors. This interaction generates corrections that are infrared finite and can be therefore neglected~\cite{Mirlin_free_2D_2024,Mirlin_free_1D_2023}. As a result, one can replace $\bar{g}[\mathcal{Q}_0]$ in the equation above by $\bar{g}[\hat{\mathcal{Q}}_{\msf{sp}}]=1$.

\subsection{Renormalization from massive modes}\label{sec:app-nlsm-mm}

Let us now examine the renormalization effects arising from the massive modes generated by the rotation $\hat{\mathcal{R}}_{\Phi}=e^{i\Phi\otimes \hat{\tau}^2/2}$, which acts on both $\hat{\bar{q}}$ and $\hat{Q}$, as well as the relative rotation $\hat{\tilde{\mathcal{R}}}$ between them. 
Specifically, we now include the massive modes generated by $\hat{\mathcal{R}}_{\Phi}$ and $\hat{\tilde{\mathcal{R}}}$, and consider fluctuations described by
\begin{equation}
\begin{aligned}
 \hat{Q}_{tt',\rb}= \hat{\mathcal{R}}_{t,\rb}\hat{\tilde{\mathcal{R}}}_{ t,\rb}\hat{\tau}^3\hat{\tilde{\mathcal{R}}}_{t,\rb}^{-1}\hat{\mathcal{R}}_{t,\rb}^{-1}\delta_{tt'},
      \quad
      \hat{\bar{q}}_{t,\rb}=i\gm {\hat{\mathcal{R}}}_{t,\rb}\hat{\tau}^3{\hat{\mathcal{R}}}_{t,\rb}^{-1},
\end{aligned}
\label{eqn:Qqnew}
\end{equation}
with the  unitary rotation $\hat{\mathcal{R}}$ now given by $\hat{\mathcal{R}}=\hat{\mathcal{R}}_{\Theta}\hat{\mathcal{R}}_{\Phi}\hat{\mathcal{R}}_{\theta}\hat{\mathcal{R}}_{\phi}$. The relative rotation $\hat{\tilde{\mathcal{R}}}$ takes the same form as $\hat{\mathcal{R}}$ but with $\hat{\Theta}$, $\hat{\Phi}$, $\theta$, and $\phi$ replaced by $\hat{\tilde{\Theta}}$, $\hat{\tilde{\Phi}}$, $\tilde{\theta}$, and $\tilde{\phi}$, respectively.

Substituting this parameterization into Eq.~\ref{eqn:SQq}, we find the quadratic in $\hat{q}$ terms become
\begin{equation}
\begin{aligned}
    &-\frac{1}{2g^2\alpha_{f}}\int_{t,\mathbf{r}}
    \left[\Tr\left[(\hat{q}_{t,\mathbf{r}}\hat{\tau}^1)^2\right]-f \Tr^2{(\hat{q}_{t,\mathbf{r}}\hat{\tau}^1)}\right]
    \\
    &=\left(\frac{\Lambda_k}{2\pi}\right)^d\frac{\gm}{4}
    \int_{t,t^\prime,\mathbf{r}}
    \left\lbrace
    \Tr\left[\left(\hat{\mathcal{Q}}_0\left(\cos\hat{\Phi}\hat{\tau}^1-\sin\hat{\Phi}\hat{\tau}^3\right)\right)^2\right]
    \right.
    \\
    &\left.\qquad\qquad \qquad
    -
    f\Tr^2\left[\hat{\mathcal{Q}}_0\left(\cos\hat{\Phi}\hat{\tau}^1-\sin\hat{\Phi}\hat{\tau}^3\right)\right]
    \right\rbrace.
\end{aligned}
\label{eq:SMM-0}
\end{equation}
Retaining terms up to quadratic order, this yields the bare action for the massive mode $\hat{\Phi}$, 
\begin{align}
    iS_{\Phi}=\left(\frac{\Lambda_k}{2\pi}\right)^d\gm
    \tr_{R}(\hat{\Phi}^2).
\end{align}

 The $\Tr\ln$ term in Eq.~\ref{eqn:SQq}, after substituting Eq.~\ref{eqn:Qqnew}, becomes
\begin{equation}\label{eq:SMM-1}
\begin{aligned}
    &\Tr\ln\left[\hat{G}_0^{-1}+i\gm \hat{\mathcal{R}} \hat{\tau}^3 \hat{\mathcal{R}}^{-1}+i\gd \hat{\mathcal{R}}\hat{\tilde{\mathcal{R}}} \hat{\tau}^3 \hat{\tilde{\mathcal{R}}}^{-1}\hat{\mathcal{R}}^{-1}
    \right]
    \\
    =&\Tr\ln\left[1+\hat{G}_{\msf{sp}}\hat{\mathcal{R}}^{-1}
    \left[\hat{G}_0^{-1},\hat{\mathcal{R}}\right]
    +i\gd \hat{G}_{\msf{sp}} \hat{\tilde{\mathcal{R}}} \left[\hat{\tau}^3, \hat{\tilde{\mathcal{R}}}^{-1}\right]
    \right].
\end{aligned}
\end{equation}
Expanding the equation above, the leading order contribution contains $i\Tr\left[\hat{G}_{\msf{sp}} \hat{\mathcal{R}}^{-1} \partial_t \hat{\mathcal{R}}\right]$ as before. Apart from the usual terms in Eq.~\ref{eq:SGM-3}, the inclusion of $\hat{\mathcal{R}}_{\Phi}$ in  $\hat{\mathcal{R}}$ further generates an interaction between the massive mode $\hat{\Phi}$ and the soft modes $\hat{\mathcal{Q}}_0$ and $\hat{U}$: 
\begin{align}
\begin{aligned}
    &
     iS_{c1}=\frac{1}{2}\left(\frac{\Lambda_k}{2\pi}\right)^d
     \Tr\left[(\hat{\mathcal{R}}_{\Phi}\hat{\mathcal{Q}}_0 \hat{\mathcal{R}}_{\Phi}^{-1} -\hat{\mathcal{Q}}_0)\hat{\mathcal{R}}_{\Theta}^{-1} \partial_t \hat{\mathcal{R}}_{\Theta}\right]
     \\
     &=
     -\frac{1}{4}\left(\frac{\Lambda_k}{2\pi}\right)^d
     \tr_{K}\left[\hat{\mathcal{Q}}_0\hat{\tau}^3\right]\tr_{R}\left[\hat{\Phi} \hat{U}^{-\frac{1}{2}}\partial_t \hat{U} \hat{U}^{-\frac{1}{2}}\right].
\end{aligned}
\end{align}
As before, we  set $\hat{\mathcal{Q}}_0$ to its saddle point in the prefactor, neglecting higher-order corrections from its coupling to $\hat{U}$ and $\hat{\Phi}$.

The $\Tr\ln$ expansion also gives rise to the bare action for the massive modes associated with $\hat{\tilde{\mathcal{R}}}$,
\begin{align}
    \begin{aligned}
    &iS_{\tilde{\Phi}}\!=\!\!
    i\gd
    \Tr\left[\hat{G}_{\msf{sp}}\hat{\tilde{\mathcal{R}}}  \left[\hat{\tau}^3,\hat{\tilde{\mathcal{R}}}^{-1}\right]\right]
    +
    \frac{\gamma_{\msf{D}}^2}{2}
        \Tr\left[ \left(\hat{G}_{\msf{sp}}\hat{\tilde{\mathcal{R}}}  \left[\hat{\tau}^3, \hat{\tilde{\mathcal{R}}}^{-1}\right]\right)^2\right]
        \\
     &\approx
     -\left(\frac{\Lambda_k}{2\pi}\right)^d
     \frac{\gd\gm}{2\tilde{\gamma}}
     \left(\tr_R[\hat{\tilde{\Phi}}^2]+\tr_R[\hat{\tilde{\Theta}}^2]+R\tilde{\phi}^2+R\tilde{\theta}^2\right),
    \end{aligned}
\end{align}
and their leading order interaction with the massless mode $\hat{U}$,
\begin{align}
    \begin{aligned}
    iS_{c2}=&\gd\Tr\left\lbrace
    \hat{G}_{\msf{sp}} \left(\hat{\mathcal{R}}^{-1} \partial_t \hat{\mathcal{R}}\right)\hat{G}_{\msf{sp}}\hat{\tilde{\mathcal{R}}} \left[\hat{\tau}^3, \hat{\tilde{\mathcal{R}}}^{-1}\right]
    \right\rbrace
    \\
     =&
     -\left(\frac{\Lambda_k}{2\pi}\right)^d
     \frac{\gd}{2\tilde{\gamma}}
     \tr_{R}\left[\hat{\tilde{\Phi}} \hat{U}^{-\frac{1}{2}}\partial_t \hat{U} \hat{U}^{-\frac{1}{2}}\right].
    \end{aligned}
\end{align}

The action $S_{mm}=S_{\Phi}+S_{\tilde{\Phi}}+S_{c1}+S_{c2}$, describing the massive modes and their coupling to the massless modes, can be used to integrate out the massive modes, yielding an effective action for the Goldstone mode $\hat{U}$. The resulting action has the same form as Eq.~\ref{eq:SGM-F1}, second line, except that the time derivative term $\left(\partial_t \hat{U}^{-1} \partial_t \hat{U}\right)$ is renormalized by a prefactor $z=\tilde{\gamma}/{2\gm}$, consistent with the Gaussian result in Sec.~\ref{Sec:Gau}.

\subsection{Coupling to the source field}\label{sec:app-nlsm-s}

The final step in the derivation of the NL$\sigma$M is to incorporate the coupling to the source field $\hat{J}$, required for the computation of the averaged-density correlation function and particle-number cumulants. The source field enters through the $\Tr\ln$ term in Eq.~\ref{eq:SMM-1}:
\begin{equation}\label{eq:SMM-J}
\begin{aligned}
    &\Tr\ln\left[\hat{G}_0^{-1}+i\gm \hat{\mathcal{R}} \hat{\tau}^3 \hat{\mathcal{R}}^{-1}+i\gd \hat{\mathcal{R}}\hat{\tilde{\mathcal{R}}} \hat{\tau}^3 \hat{\tilde{\mathcal{R}}}^{-1}\hat{\mathcal{R}}^{-1}
    -
    \hat{\mathcal{J}}
    \right]
    \\
    =&\Tr\!\ln\!\!\left[
    1
    +\hat{G}_{\msf{sp}}\!
    \left(
    \hat{\mathcal{R}}^{-1}\!\!\!
    \left[\hat{G}_0^{-1},\hat{\mathcal{R}}\right]
    +i\gd \hat{\tilde{\mathcal{R}}} \! \left[\hat{\tau}^3, \hat{\tilde{\mathcal{R}}}^{-1}\right]
    -\!
    \hat{\mathcal{R}}^{-1}
    \hat{\mathcal{J}}
    \hat{\mathcal{R}}
    \right)
    \right].
\end{aligned}
\end{equation}

Expanding the $\Tr\ln$ to first order yields one source-field-dependent term, in addition to those discussed above, and it can be expressed as
\begin{equation}\label{eq:SMM-J1}
\begin{aligned}
    &-\Tr\left[
    \hat{G}_{\msf{sp}} \hat{\mathcal{R}}^{-1}
    \hat{\mathcal{J}}
    \hat{\mathcal{R}}
    \right]
    =iS_{JQ_0}+iS_{J\Phi},
    \\
    &iS_{JQ_0}
    =
    \frac{i}{2}\left(\frac{\Lambda_k}{2\pi}\right)^d
    R\tr_{K}\left[\hat{\mathcal{J}}^{\msf{q}(s)}\hat{\tau}^1\hat{\mathcal{Q}}_0\right],
    \\
    &iS_{J\Phi}
    =
    -\frac{i}{4}\left(\frac{\Lambda_k}{2\pi}\right)^d
    \tr_{K}\left[\hat{\mathcal{Q}}_0\hat{\tau}^3\right]
    \tr_{R}\left[\hat{\Phi} 
    \hat{\mathcal{J}}^{U}
    \right],
\end{aligned}
\end{equation}
where 
\begin{align}
    \hat{\mathcal{J}}^{U}\equiv\left(\hat{U}^{-\frac{1}{2}}\hat{\mathcal{J}}^{\msf{q}(a)} \hat{U}^{\frac{1}{2}}
    +\hat{U}^{\frac{1}{2}}\hat{\mathcal{J}}^{\msf{q}(a)} \hat{U}^{-\frac{1}{2}}\right).
\end{align}
At second order, we have
\begin{equation}\label{eq:SMM-J2}
\begin{aligned}
    iS_{J\tilde{\Phi}}
    &=i\gd\Tr\left[
    \hat{G}_{\msf{sp}}
    \hat{\mathcal{R}}^{-1}
    \hat{\mathcal{J}}
    \hat{\mathcal{R}}
    \hat{G}_{\msf{sp}}
    \hat{\tilde{\mathcal{R}}} \! \left[\hat{\tau}^3, \hat{\tilde{\mathcal{R}}}^{-1}\right]
    \right]
 \\
     &
    \approx
    -\left(\frac{\Lambda_k}{2\pi}\right)^d
     \frac{i\gd}{2\tilde{\gamma}}
     \tr_{R}\left[\hat{\tilde{\Phi}}
    \hat{\mathcal{J}}^{U}
    \right],
\end{aligned}
\end{equation}
and
\begin{equation}\label{eq:SMM-J21}
\begin{aligned}
    iS_{JU}&=i\Tr\left[
    \hat{G}_{\msf{sp}}
    \hat{\mathcal{R}}^{-1}
    \hat{\mathcal{J}}
    \hat{\mathcal{R}}
    \hat{G}_{\msf{sp}}
    \hat{\mathcal{R}}^{-1}
    \partial_t\hat{\mathcal{R}}
    \right]
    \\
    &
    \approx
    \left(\frac{\Lambda_k}{2\pi}\right)^d
     \frac{i}{4\tilde{\gamma}}
     \tr_{R}\left[\hat{U}^{-\frac{1}{2}}\partial_t \hat{U} \hat{U}^{-\frac{1}{2}} \hat{\mathcal{J}}^{U}\right].
\end{aligned}
\end{equation}
An additional quadratic-in-$J$ term arises from the second order expansion:
\begin{equation}\label{eq:SMM-J3}
\begin{aligned}
    &iS_{J0}=-\frac{1}{2}\Tr\left[
    \left(\hat{G}_{\msf{sp}} \hat{\mathcal{R}}
    \hat{\mathcal{J}}
    \hat{\mathcal{R}}^{-1}
    \right)^2
    \right]
    \\
    &=
    -\left(\frac{\Lambda_k}{2\pi}\right)^d
    \frac{1}{2\tilde{\gamma}}
    \left\lbrace
    R\left[\hat{\mathcal{J}}^{\msf{q}(s)}\hat{\mathcal{J}}^{\msf{q}(s)}\right]
    +\frac{1}{4}
    \tr_R\left[\left(\hat{\mathcal{J}}^{U}\right)^2\right]
    \right\rbrace.
\end{aligned}
\end{equation}

As shown above, in addition to the direct coupling to the Goldstone modes $\hat{\mathcal{Q}}_0$ and $\hat{U}$ ($S_{JU}$ and $S_{JQ_0}$), 
the source field $\hat{J}$
also couples to the massive modes ${\hat{\Phi}}$ and $\hat{\tilde{\Phi}}$ ($S_{J\Phi}$ and $S_{J\tilde{\Phi}}$), which in turn interact with the massless modes at quadratic order ($S_{c1}$ and $S_{c2}$).
Integrating out these massive modes therefore renormalizes the coupling between the source field and massless modes, as well as the quadratic source field action in $S_{J0}$.
More specifically, combining $S_{J\Phi}+S_{J\tilde{\Phi}}$ with the massive mode action $S_{mm}$ and integrating out the massive modes, we obtain the effective action
\begin{align}
    \begin{aligned}
    iS_{r}
    =
    -\frac{(1-z)}{8\tilde{\gamma}}
    \tr_{R}\left[
    \left(
    \hat{U}^{-\frac{1}{2}}\partial_t \hat{U} \hat{U}^{-\frac{1}{2}} 
    +i\hat{\mathcal{J}}^{U}
    \right)^2
    \right].
    \end{aligned}
\end{align}
Here the $J$-independent term renormalizes the Goldstone action $S_{gm}$ in the replica-asymmetric sector by an overall factor $z$ in the $\partial_t$-dependent term, as explained in Sec.~\ref{sec:app-nlsm-mm}. The remaining terms provide the same renormalization factor $z$ in both the source-massless coupling ($S_{JU}$) and the $\mathcal{J}^{U}$ dependent term in the source field action $S_{J0}$.
Combining all terms and including all these renormalization effects, we arrive at the NL$\sigma$M in Eq.~\ref{eq:SGM-F2}.

\section{Numerical Method}\label{sec:app-numerics}
We simulate the monitored dynamics using quantum projection dynamics (QPD), a discrete measurement scheme that is equivalent to the continuous monitoring used in the analytical treatment. We use QPD rather than quantum state diffusion \cite{Diehl_free_1D_numerical_2021} because the projective updates admit an efficient rank-1 reorthonormalization scheme discussed in the following.

The numerical algorithm is summarized as follows,
\begin{enumerate}
    \item Start with the initial wave function $\Psi_0$, which is represented with a $N\times N/2$ matrix.
    \item At time $t$, we have the wave function $\Psi_t$, and we perform time evolution $\Psi_{t+\delta t} = e^{-iH\delta t} \Psi_t$. Here we use fixed $\delta t = 1/(N\gamma)$. The time evolution is evaluated with Chebyshev expansion (discussed in detail below), which is highly efficient and expanded to order with machine precision.
    \item Quantum projection at site $k$, which is randomly selected across the system with equal probability for all the sites. At the picked site $k$, density measurement is performed and there are two possible outcomes: $n_k=1$ (on) and $n_k=0$ (off), occurring with probability  $P_k = \left\langle n_{k}\right\rangle$  and $1-P_k$ , respectively. The wave function update can be written as $\Psi' = \Psi + u v^\dagger$ with $u$ and $v$ being $N\times 1$ and $N/2\times 1$ column vectors, respectively,
    \begin{align}
            \label{eq:uv}
        u_i  & =\begin{cases}
        \frac{\delta_{ik}}{\sqrt{\braket{n_k}}}-\frac{\sum_j \Psi_{ij}\Psi^*_{kj}}{\braket{n_k}} & \text{ on}\,,\\
        -\delta_{ik} & \text{ off}\,.
        \end{cases}\\
        v_j & =\Psi_{kj}^{*}\,.
    \end{align}
    \item Reorthonormalization of $\Psi'$.
    \item Repeat 2-4 until convergence criteria are met.
    \item Calculate physical observables.
\end{enumerate}

For projecting on, the measured site is projected to unit occupation and decouples from the other sites; for projecting off, its occupation is projected off.
The project-on operation defined in Eq.~\eqref{eq:uv} is identical to the quantum jump operation discussed in, e.g. Ref.~\cite{Diehl_free_1D_numerical_2021}, where the jump operation is performed on the correlation matrix and orthonormality is restored via SVD decomposition. Here, we instead work with the wave functions directly. For the project-on outcome, the resulting wave function remains exact column orthonormal, and no subsequent SVD or QR decomposition is required. For the project-off outcome, the reorthonormalization is performed with the rank-$1$ polar update scheme discussed below. The rank-$1$ method is much more efficient than QR or SVD, especially on modern hardware with high-bandwidth memory, e.g. GPUs.

\subsection{Chebyshev time evolution}
We employ Chebyshev expansion of the evolution operator $e^{-i H \Delta t}$ for the unitary evolution between quantum projections. The Chebyshev time evolution is numerically stable, highly accurate, and efficient for sparse Hamiltonians. The time evolution operator can be expressed as
\begin{equation}
    U\left(t\right)=e^{\frac{i\beta t}{\alpha}}\left[J_{0}\left(\frac{t}{\alpha}\right)+2\sum_{n\ge1}\left(-i\right)^{n}J_{n}\left(\frac{t}{\alpha}\right)T_{n}\left(\tilde{H}\right)\right]\,.
\end{equation}
Here the Hamiltonian is rescaled with $\tilde{H} = \alpha H + \beta$ so that the spectrum of $\tilde{H}$ falls in the range $(-1, 1)$, where Chebyshev expansion is convergent. $J_n(x)$ is the Bessel function of the first kind (not to be confused with the source field), and $T_n(x)$ is the Chebyshev polynomial of the first kind. Thus, we have
\begin{align}
    &\ket{\psi(t+\Delta t)}  
    =U\left(\Delta t\right)\ket{\psi(t)}\nonumber \\
    =& 
    e^{\frac{i\beta \Delta t}{\alpha}}
    J_{0}\left(\frac{\Delta t}{\alpha}\right)\ket{\psi_{0}(t)}
    +
    2e^{\frac{ib \Delta t}{a}}
    \sum_{n\ge1}(-i)^{n} J_{n}\left(\frac{\Delta t}{\alpha}\right)\ket{\psi_{n}(t)}\,.
\end{align}
Here the $\ket{\psi_n(t)}$ is given by
\begin{equation}
   \ket{\psi_{n}(t) }  =T_{n}\left(\tilde{H}\right)\ket{\psi(t)}\,.
\end{equation}
When the Hamiltonian is sparse, the operation can be performed very efficiently, especially on modern GPUs with high bandwidth memory. 

The $\ket{\psi_{n}}$ can be obtained iteratively with recurrence
relation,
\begin{align}
\begin{aligned}
\ket{\psi_{0}} & =\ket{\psi\left(t\right)}\,.\\
\ket{\psi_{1}} & =\tilde{H}\ket{\psi_{0}}\,,\\
\qquad & \cdots\\
\ket{\psi_{n+1}} & =2\tilde{H}\ket{\psi_{n}}-\ket{\psi_{n-1}}\,.
\end{aligned}
\end{align}
 The Chebyshev expansion of the evolution operator converge rapidly 
because, at fixed argument, the Bessel function $J_n(z)$ falls to $0$ superexponentially with $n$,
\begin{equation}
J_{\nu}\left(z\right)\sim\frac{1}{\Gamma\left(\nu+1\right)}\left(\frac{z}{2}\right)^{\nu}\,,\qquad0<\left|z\right|\ll\sqrt{\nu+1}\,.
\end{equation}
 Therefore, only a small number of Chebyshev polynomials are required
for accurate evaluation of $U\left(t\right)$. 

The upper bound of the error with cutoff at order $M$ can be estimated
via
\begin{align}
\Delta & =\left|\braket{\psi_\mathsf{exact}\left(\Delta t\right)|\psi_\mathsf{exact}\left(\Delta t\right)-\psi_\mathsf{Cheb}\left(\Delta t\right)}\right|\nonumber \\
 & =\left|2\sum_{n=M+1}^{\infty} (-i)^n J_{n}\left(\frac{t}{a}\right)\braket{\psi_\mathsf{exact}\left(t\right)|\psi_{n}}\right|\nonumber \\
 & <2\sum_{n=M+1}^{\infty}\left|J_{n}\left(\frac{t}{a}\right)\right|\nonumber \\
 & \approx\sqrt{\frac{2}{M\pi}}\left(\frac{t}{2a}\right)^{M}e^{-M\left(\log M-1\right)}\,.
\end{align}
The error vanishes with $M$ faster than exponential function, and machine precision can be achieved with small $M$ when the measurement time step is small.  

\subsection{Rank-$1$ polar reorthonormalization}

The orthonormalization of the wave function is usually restored either with QR decomposition of wave function $\Psi$ or SVD of the correlation matrix $D$. The projection operation discussed in this manuscript is effective rank-$1$ and we can then perform the rank-$1$ polar update to restore orthonormality. The rank-$1$ update gives the same occupied subspace as the full SVD, at much lower computational cost. The rank-$1$ polar update is as follows,
\begin{enumerate}
\item Project the perturbation
\begin{enumerate}
\item $\Psi'=\Psi+uv^{\dagger}$ and $w=\Psi^{\dagger}u$
\end{enumerate}
\item Build $2D$ orthonormal basis
\begin{enumerate}
\item $v_{1}=v/\left|\left|v\right|\right|$
\item $w_{\perp}=w-\left(v_{1}^{\dagger}w\right)v_{1}$
\item $v_{2}=w_{\perp}/\left|\left|w\right|\right|$
\item Basis matrix $V=\left[v_{1}\quad v_{2}\right]$
\item $\alpha_{1}=v_{1}^{\dagger}w$ and $\alpha_{2}=\left|\left|w_{\perp}\right|\right|$
\end{enumerate}
\item Formulate the metric tensor $Y$ 
\[
Y=\left[\begin{array}{cc}
Y_{11} & Y_{12}\\
Y_{21} & Y_{22}
\end{array}\right]\,,
\]
with 
\begin{align}
\begin{aligned}
Y_{11} & =1+2\Re\left(\alpha_{1}\left|\left|v\right|\right|\right)+\left|\left|u\right|\right|^{2}\left|\left|v\right|\right|^{2}\,,\\
Y_{12} & =\alpha_{2}\left|\left|v\right|\right|\,,\\
Y_{21} & =\alpha_{2}\left|\left|v\right|\right|\,,\\
Y_{22} & =1\,.
\end{aligned}
\end{align}
\item Analytical inverse square root $\left(Z=Y^{-\frac{1}{2}}\right)$
\[
Z=\frac{1}{s\cdot t}\left[\begin{array}{cc}
Y_{22}+s & -Y_{12}\\
-Y_{21} & Y_{11}+s
\end{array}\right]\,,
\]
with 
\begin{align}
s & =\sqrt{\det Y}\,,\\
t & =\sqrt{\mathsf{Tr}Y+2s}\,.
\end{align}
\item Rotation-aware matrix update
\[
\Psi'_{\mathsf{unitary}}=\Psi+\left(\Psi V\Delta Z+W\right)V^{\dagger}\,,
\]
with $\Delta Z=Z-I_{2}$ and $W=u\left[\left|\left|v\right|\right|Z_{11}\quad\left|\left|v\right|\right|Z_{12}\right]$.
\end{enumerate}

For the project-off case, the rank-1 polar update can be further simplified as,
\begin{equation}
\Psi'_\mathsf{unitary}=\Psi+u_\mathsf{off}v^\dagger ,
\end{equation}
where
\begin{equation}
u_\mathsf{off}
=
\frac{(1 - \braket{n_k})^{-1/2}-1}{\braket{n_k}}\,
\Psi v
-
\frac{|k\rangle}{\sqrt{1-\braket{n_k}}} .
\end{equation}
with $\braket{n_k}$ being the fermion density at site $k$ before projection. 
It can be shown that $\Psi'_\mathsf{unitary}$ is column orthonormal and no further QR or SVD decomposition is required. In the numerical simulation, we periodically (every $100$ projections) re-orthonormalize the state with Newton-Schulz method to reduce the accumulating numerical errors.

\setcounter{section}{0}

\bibliography{references_temp}

\end{document}